\documentclass[a4paper,superscriptaddress, 11pt,twocolumn]{revtex4-1}
\usepackage{amsmath}
\usepackage{ulem}
\usepackage{braket}
\usepackage{ragged2e}
\usepackage{graphicx}
\usepackage{epstopdf}

\begin{document}

\title{Secure communication based on sensing of undetected photons}

\author{$J.$ $Sternberg$}
\affiliation{ Laboratoire Mat\'eriaux et Ph\'enom\`enes Quantiques (MPQ), Universit\'e Paris Cité, CNRS-UMR 7162, Paris 75013, France}

\author{$J.$ $Voisin$}
\affiliation{ Laboratoire Mat\'eriaux et Ph\'enom\`enes Quantiques (MPQ), Universit\'e Paris Cité, CNRS-UMR 7162, Paris 75013, France}

\author{$C.$ $Roux$}
\affiliation{ Laboratoire Mat\'eriaux et Ph\'enom\`enes Quantiques (MPQ), Universit\'e Paris Cité, CNRS-UMR 7162, Paris 75013, France}

\author{$Y.$ $Chassagneux$}
\affiliation{ Laboratoire de Physique de l’ENS, Université PSL, CNRS, Sorbonne Université,Université Paris Cité, Paris, France}

\author{$M.I.$ $Amanti^{*}$}

\affiliation{ Laboratoire Mat\'eriaux et Ph\'enom\`enes Quantiques (MPQ), Universit\'e Paris Cité, CNRS-UMR 7162, Paris 75013, France}

\begin{abstract}
In this paper, we introduce a secure optical communication protocol that harnesses quantum correlation within entangled photon pairs. A message written by acting on one of the photons can be read by exclusive measurements of the other photon of the pair. In this scheme a bright, meaningless optical beam hides the message rendering it inaccessible to potential eavesdroppers.  Unlike traditional methods our approach only affects unauthorized users, fundamentally limiting their access to the communication channel. We demonstrate the effectiveness of our protocol by achieving secure communication through both amplitude and phase modulation. We successfully employ this technique for the secure transfer of an image. We demonstrate data exchange speed of up to 8 bits per second, along with the corresponding eye diagrams.
\end{abstract}


\maketitle
Data transmission is essential in today's information age. Photons are ideal carriers for communication because of their speed and low sensitivity to interlink interference.  Nowadays, a global fibre-optic network, together with free-space optical links, provides high-bandwidth optical communications. In situations where sensitive data are involved, secure channels are imperative to prevent information leakage.
Quantum physics has proven to be a robust tool for implementing secure communication schemes. Current protocols use the quantum no-cloning theorem \cite{Wootters1982}, which states the impossibility of detecting a photon, extracting all the quantum information it contains, and subsequently transmitting another photon as an exact quantum copy of the original. An user wishing to extract information from the communication photons stream must perform some form of measurement on it. If data are encoded using quantum-mechanical means, any potential eavesdropper will be forced to reveal his presence, due to the invasive nature of the measurements. This represents the fundamental principle behind quantum cryptography. Numerous protocols based on single-photon sources or entangled pairs have been demonstrated \cite{Bennett1992, 1982PhLA...92..271D}. Quantum key distribution (QKD), which involves the remote delivery of a secret key through an insecure channel using quantum mechanics, has evolved from a theoretical interest to a thriving industry. Nowadays, both metropolitan and satellite-based QKD networks have been demonstrated \cite{Stucki_2011,Yang:21,Chen2021}. Furthermore, the security limitations of QKD, stemming from the assumption that quantum devices behave in line with given mathematical models, have been surpassed in device-independent schemes \cite{Zapatero2023}. These protocols do not necessitate the characterization of any device's internal functioning, but they still rely on a set of assumptions concerning the sharing of a secret key and the generation of randomness.
In the present work we propose an alternative method for secure optical communication. This approach is based on the use of quantum correlation within entangled photon pairs as a means to hide a message by drowning it in a bright meaningless optical beam.  
In general terms, communication between two users involves one party writing a message and sending it over a chosen link to the second user, who can then read it. Each form of reading is based on a measurement of some physical quantity transmitted over the selected link. If the link is a photon stream, reading means the conversion of the optical signal into an electrical signal in general, or any other physical quantity. If an unauthorized party wants to access the message, he has to intercept part or all of the optical signal and try to read it. This detection process is a statistical phenomenon that has a certain probability of success and is limited by shot noise in its performance, according to Poissonian statistics \cite{loudon_quantum_2000,mandel_wolf_1995}.
Even with an unrealistically high level of efficiency, the eavesdropper would still face the statistical nature inherent in the optical carrier when trying to intercept the message. In situations involving a coherent source like a laser as jamming beam, a Poissonian statistic reemerges in the measurement process, and the associated fluctuation establishes the ultimate limit of sensitivity in deciphering the message. A weak beam of light carrying an optical message can be concealed by a powerful laser, analogous to a message written in white ink on a white page. Nevertheless, the question arises: How can we ensure accessibility of the message to the intended recipient? Quantum mechanics plays a crucial role in this context. In this work we will exploit the quantum correlation present in entangled pairs of photons. One of the photons will travel between the two users, while the other remains on the receiver's location and it is thus insensitive to external attacks. The photon deployed for transmission is used to write the message that is hidden by an intense stream of photons emitted from a coherent source, rendering it inaccessible. The other photon in the pair, which is not impacted by the jamming laser, is utilized to read the message via correlations in quantum pairs. The technique of concealing a message in a carrier signal is widely used and exploits the seemingly innocuous nature of the transmitted data \cite{e21040355,cachin_informationtheoretic_2004,wang2008perfectly}. This approach is especially effective when the volume of publicly available information is so vast that checking everything for potential secret message is prohibitively time consuming. The quantum version of this scheme has been demonstrated and an exhaustive review is presented in \cite{app122010294}. In these works secret data are embedded into a cover message and are encoded using qubits rather than classical bits. Security relies on the complexity of the encoding of the hidden message, as in its classical equivalent, or in the fundamental limitation of quantum non cloning theorem, as for the more widely developed QKD. In all these protocols the meaningless message covers the reading of the true message, while in the scheme proposed in this work, only the unwanted user suffers from the hiding source. The ability of the eavesdropper to have access to the communication is fundamentally limited by the jamming light, regardless of the properties of the quantum sources establishing the link. Meanwhile, correlations in photon pairs have proven to be an effective means of mitigating background noise in coincidence measurements within quantum illumination schemes \cite{gregory_imaging_2020,lloyd_enhanced_2008,clark_special_2021,moreau_demonstrating_2017,moreau_imaging_2019}. Recent developments include heralded measurements in the presence of noise with the same spectral characteristics as the signal to be detected \cite{johnson_hiding_2023}. However this approach applied to communication encounters a drawback, as the noise directly impacts the detection of the data to be measured. In contrast, in the presented protocol, noise only affects the undetected photon. 

\section{Results} 
\subsection{Secure communication protocol}
\begin{figure} 
\includegraphics[trim={0 0 0 0}, scale=0.4]{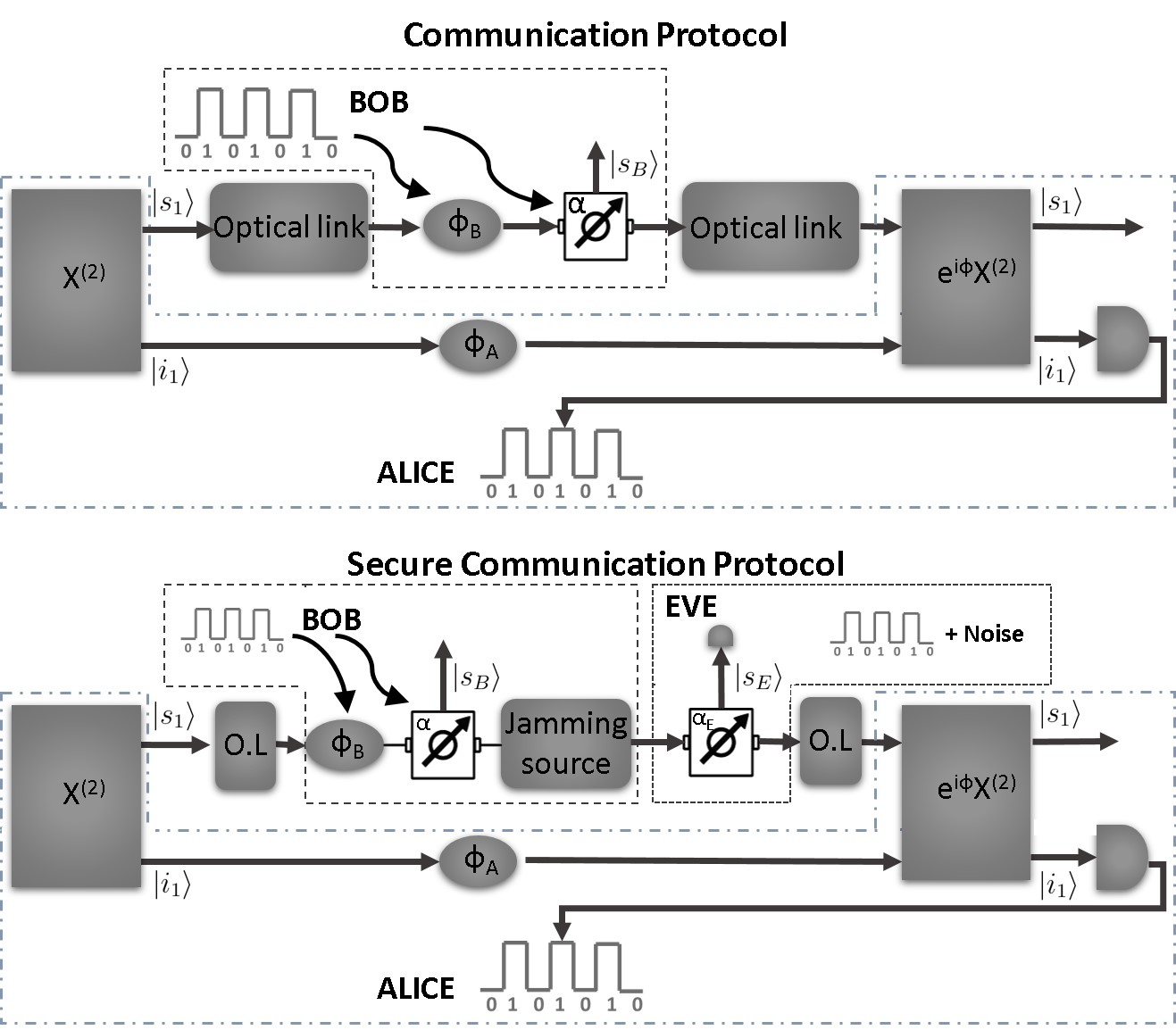}
\centering 
\caption{Schematic representation of the communication protocol based on sensing of undetected photons (Top Panel). The two users are Alice and Bob. A quantum source ($\chi^{2}$) generates two photons on two different modes. $\ket{s_1}$ goes to Bob through an optical link while $\ket{i_1}$ stays at Alice's location. Bob writes the message on the phase ($\phi_B$) or on the amplitude ($\alpha^2$) before sending the message back to Alice. At her location a second quantum source is present ($e^{i\phi}\chi^{2}$). The bottom panel shows the scheme of the secure communication. Eve is the eavesdropper who steals $\alpha_{Eve}^2$ of the mode $\ket{s_1}$. OL is the optical link. A jamming source is added by Bob to hide the message.}
\label{figProtocol}
\end{figure}

In this work, we propose a secure optical communication protocol utilizing sources of entangled photon pairs (see Figure \ref{figProtocol}, Top panel).
A given user, Alice, owns a photon pair source ($\chi^{(2)}$) generating an idler and a signal in the quantum state
$\ket{\psi}=\ket{i{_1}}\ket{s{_1}}$, where $i_1/s_1$ is the mode of the idler/signal photon respectively. Alice establishes an optical link with a second user, Bob, along the optical mode $\ket{s_1}$. Bob extends the link with a connection that goes back to Alice. The mode $\ket{i_1}$, on the other side, stays at Alice's location, where a second source of photon pairs is present ($e^{i\phi}\chi^{(2)}$). This is an exact replica of the first source and it emits photon pairs on the same modes $s_1$ and $i_1$ with a specific phase $\phi$ respect to the first source. If the emission from the two sources is coherent, and the probability of simultaneous emission from the two sources is negligible, the quantum state, once the communication link is established, is :
\begin{equation}
\begin{aligned}
\ket{\psi_{link}} = \ket{vacuum}+\\
 \sqrt{N_{Quantum}}\ket{i_{1}}\ket{s_{1}}(1+e^{i(\phi-\phi_A-\phi_B)})
\end{aligned}
\label{eq:initalstate1}
\end{equation}
where the phase terms $\phi_B$ and $\phi_A$, are related to the delay accumulated along the optical paths $\ket{s_{1}}$ and $\ket{i_{1}}$ respectively (See Supplementary Note 1) and $\ket{vacuum}$ is the vacuum state. $N_{Quantum}$ is the number of pairs generated per second, per source. We work here in the low gain regime where high order terms including more than two photons generation are neglected.
To encode the message, Bob can manipulate the quantum state in the region between the two optical links by adjusting the phase and the amplitude of the signal mode that he sends back to Alice. In the first scenario, he can vary the optical path length along the mode $s_1$, and in the second case, he can redirect a portion of the mode to an alternate path, referred to as $s_B$.
The message is read by Alice by measuring $exclusively$ the photon that she always kept at her place. The detection rate at Alice's position $R_{A}$ is:
\begin{equation}
R_{A}=2\eta_{det}N_{Quantum}[1+ \sqrt{1-\alpha^2} cos(\phi-\phi_B-\phi_A)],
\label{eq:rateAlice}
\end{equation}
where $\alpha^2$ is the probability of deviation of the mode $s_1$ into $s_B$ at Bob's place and $\eta_{det}$ is the detection efficiency, including propagation losses and detector quality (See Supplementary Note 1). 
From equation \ref{eq:rateAlice}, we see that measurements at Alice's place are affected by Bob's choice of the phase $\phi_B$ and on his action on the transmission amplitude (on the $\alpha$ parameter). 
A binary communication system can be established between the two users, thanks to quantum correlations in entangled photon pairs.
Bob writes a message based on the phase, the amplitude (or both) of the quantum state, and Alice reads it on the idler. 
Conventional protocols available in classical communication can be directly extended to the presented protocol.
However, this link is highly vulnerable to possible eavesdroppers.
The presence of a third user in the scheme can be represented as a loss channel on the signal photon's spatial mode (refer to Figure \ref{figProtocol}, bottom panel). We model it as a beam splitter that deviates part of the mode $s_{1}$ into another mode,  $s_{E}$,  with a probability $\alpha_E^2$ (See Supplementary Note 2). 
Users can detect the presence of an eavesdropper by observing the visibility of the communication's 0 and 1 levels, upon which it relies. Meanwhile, the intruder can easily access the message addressed to Alice by intercepting the signal photons.
To secure the communication protocol against external attacks, we will now take another step.
The proposed approach relies on the inclusion of a strong light source that generates photons in the same mode $s_1$ and that are not correlated with the entangled photons. They will act as same sort of strong glare, blinding Eve. In this scheme, detection at Alice's location is unaffected by the jamming source since she only measures the idler photons. These photons never left Alice's location, they remain secured and they are not correlated with the jamming source. On the other hand the eavesdropper's detection is strongly affected by the additional light since he measures the photon population in the $s_E$ mode. This contribution comes, certainly from the photon encoded with the message but also from the strong jamming beam. In the case of poor detection efficiency, $\eta_{det} \ll 1$, the statistics of the measurement is Poissonian irrespective of the underlying statistics of the classical jamming source. In the case of higher detection efficiency the statistics is dominated by the one of the incident light. If a laser is employed as jamming source, a Poissonian statistic is retrieved. The condition to achieve a secure communication link between Alice and Bob is that the noise associated with the detection of the jamming light and of the message is stronger than the message amplitude itself \cite{Taylor1982,FOX_Book}. 
The signal to noise ratio of the communication detection by Eve is $SNR=\sqrt{C_{comm}}/\sqrt{1+2\frac{N_{Class}}{N_{Quantum}}}$ and the security condition is (See Supplementary Note 2):
\begin{equation}
\begin{aligned}
\frac{N_{Class}}{N_{Quantum}}>\frac{C_{comm}-1}{2}\approx\frac{C_{comm}}{2}=\\
\frac{\eta_{det_E}\alpha_E^2{N_{Quantum}} T_{meas}}{2},
\end{aligned}
\label{eq:fundamental}
\end{equation}
where $N_{Class}$ is the number of classical photons per second in the $s_1$ mode, $C_{comm}$ are the counts associated with the communication (difference between the counts of the upper and lower communication level) and $T_{meas}$ is the measurement time. This last parameter will determine the maximum speed for the data exchange.

This is the only requirement of the presented communication protocol to ensure the link's integrity. Even if Eve intercepts a big part of the message ($\alpha_E^2\approx1$) and  with perfect detection efficiency ($\eta_{det_E}$=1), the condition for achieving a secure communication regime is easily accessible through available technologies. If a coherent light source is employed as jamming signal, non linear interaction in the second crystal generating idler photons by difference frequency generation have to be limited in order to avoid their detection at Alice's place. It can be noticed that the use of an incoherent source would relax this condition. A detailed study of this regime is beyond the scope of the present work. In Supplementary Note 3 we show that the experiments presented in this work are outside the regime of difference frequency generation.
We would like to emphasize that sub Poissonian detection can be achieved only in the case of suitable light sources and in the high efficiency detection scheme. For this reason, the choice of the statistic of the light source acting as a jamming source and its intensity guarantee the security of the message transfer between Alice and Bob. Similarly, all photon measurements conducted through balanced detection are limited by shot noise (Poissonian noise) and would fail to reveal any concealed information within the high-intensity laser beam \cite{FOX_Book}.
The secure protocol presented in this work could be extended to classical correlation based protocols where a seed beam is introduced in the signal path and a classical detection is performed at the idler position \cite{shapiro_classical_2015,cardoso_classical_2018}. In this scheme, the existence of the intense seed beam traveling from Bob to Alice, carrying the information to be shared, makes the task of concealing its presence more challenging compared to the low gain regime described here. 

\subsection{Experimental demonstration}
\begin{figure*} 
\centering 
\includegraphics[trim={0cm 0 0cm 0}, width=0.8\textwidth]{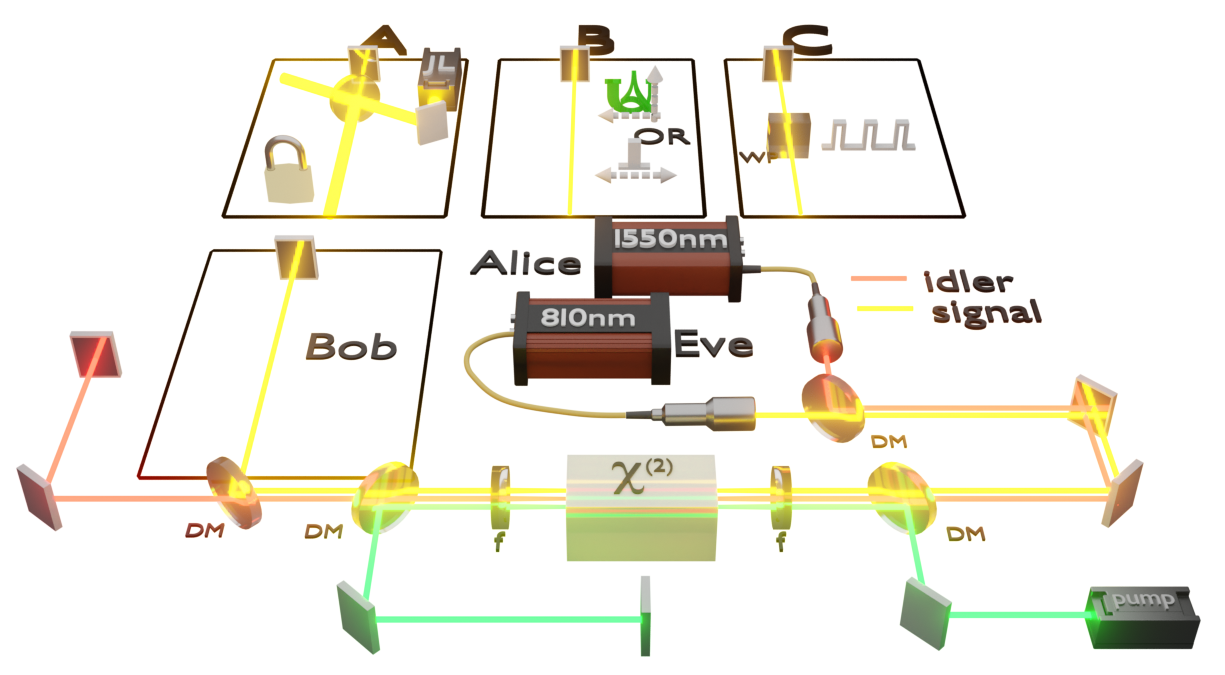}
\caption{The experimental setup comprises a second-order nonlinear crystal ($\chi^{(2)}$) excited by a continuous-wave laser operating at $\lambda$=532 nm with a power of 8 mW. Two lenses with a focal length of 5 cm (f) are employed to both focus the pump laser and collect the generated photon pairs. Dichroic mirrors (DM) are utilized to separate the pump from the photon pairs and to distinguish between the 1550 nm and 810 nm photons. Single-mode fibers are employed to couple the 1550 nm and 810 nm photons to their respective detectors. The 810 nm detector emulates the role of a potential eavesdropper in the communication channel. The three insets depict different configurations of the communication. Panel A illustrates the scenario where the jamming source overlaps with the signal mode, while Panels B and C showcase amplitude and phase-based schemes, respectively.}
\label{figsetup}
\end{figure*}
In this section we present the experimental demonstration of the secure communication protocol (See Figure \ref{figsetup}). In order to generate the quantum state used to establish the optical link (See equation \ref{eq:initalstate1}), we rely on spontaneous parametric down conversion. A 532 nm continuous wave laser acts as a pump for a second order non linear process in a PPLN crystal with a periodicity suitable for the spontaneous generation of idler/signal pairs at 1550nm/810nm. In all the presented measurements the pump power measured at the PPLN crystal is 8 mW. After a first pass in the crystal, the pump travels a given distance (25 cm) and it is sent back to the crystal along the same spatial mode as the outgoing one. This scheme mimics the sources presented in Figure \ref{figProtocol}. This scheme is commonly used in experiments based on sensing of undetected photons {\cite{PhysRevLett.67.318,PhysRevA.44.4614, Lemos2014,Kutaseaaz8065}. Alice is responsible for retaining the crystal and the 1550nm photon, whilst Bob manipulates the spatial mode of the photon at 810 nm which he shares with Alice. Two identical lenses of focal length 5 cm are used to focus the pump beam in the crystal at both passes and to collect the generated photon pairs. A first dichroic mirror separates the pump from the photon pairs and a second one splits the signal and the idler along two different paths. A planar mirror sends the optical modes back to the crystal. In the case of a perfect superposition of the mode on the first pass and second pass in the crystal, the state of equation \ref{eq:initalstate1} is retrieved and the communication protocol can be employed as described above. The detection is performed using single mode fiber coupled superconducting detectors for the idler and fiber coupled silicon single photon detectors for the signal. Idler photons are detected after being filtered with a band pass filter centered at 1550 nm (10 nm band pass) while the signal ones experience multiple filtering stages to remove the residual pump. As in classical optical links, losses determine the ultimate performances of the protocol since they govern the visibility of the transmitted message (See Supplementary Note 4).
In order to secure the communication a continuous wave laser emitting at 810 nm (10 nm bandwidth) is overlapped with the spatial mode $s_1$ (Figure 2 A). We employ a glass plate to reflect part of the laser beam on the $s_1$ mode, the remaining intensity is waisted on an useless path (See Figure \ref{figsetup} A). 
Bob encodes the message by acting on the amplitude of the signal sent back to Alice (See Figure \ref{figsetup} B) or on the phase $\phi_{B}$ of the state, (See Figure \ref{figsetup} C), in agreement with the protocol presented in the previous section.

\begin{figure*} 
\includegraphics[trim={1cm 0 0 0}, scale=0.5]{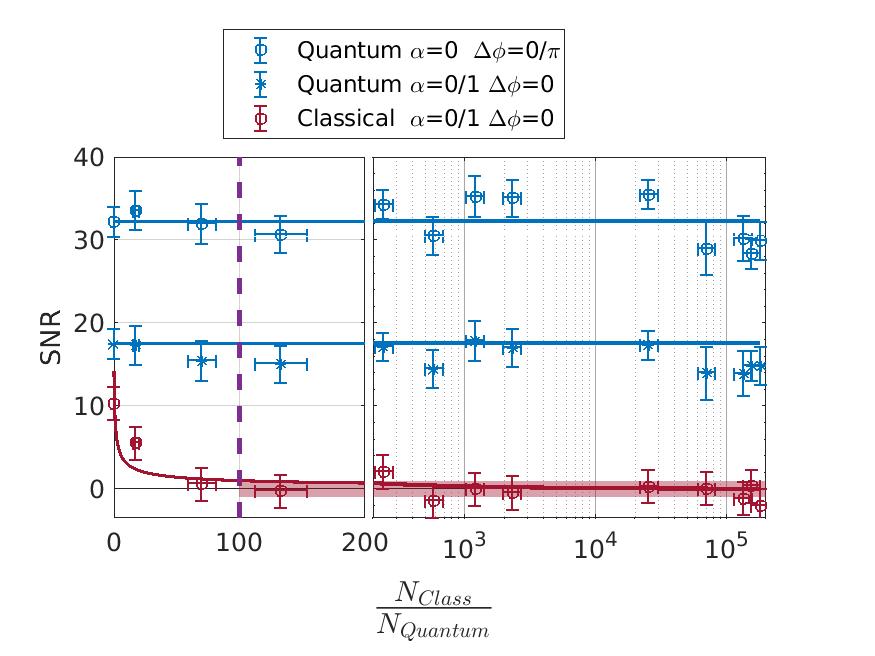}
\caption{Signal to noise ratio of the communication protocol. Experimental data correspond to measurements on the same photon used to write the message, the signal, (Classical-Eve) and on the other photon of the pair, the idler (Quantum-Alice). Data are reported as function of $\frac{N_{Class}}{N_{Quantum}}$(See Methods). Data correspond to amplitude based communication ($\alpha=0/1$) and phase based ($\Delta\phi=0/\pi$). The vertical dashed line corresponds to the condition SNR=1 and the shaded area to the secure communication regime where $-1\le SNR \le 1 .$}
\label{figSNR}
\end{figure*}
In Figure \ref{figSNR} we present the experimental dependence of the signal to noise ratio (SNR) of the communication protocol (See methods) as function of $\frac{N_{Class}}{N_{Quantum}}$, where $N_{Quantum}$ is the number of pairs generated per second and $N_{Class}$ is the number of photon per second of the jamming source that are in the same mode of the quantum photons. Results labeled as classical correspond to counts measured on the signal detector: Bob sends a message to Alice on the signal path and measurements are performed on the same path. Experimental points labeled as quantum correspond to measurements on the idler path, Bob writes on the signal and Alice measures on the idler. Classical experimental data are compared to the theoretical model (full red line), where the value $C_{Comm}$ has been extracted from the measurements at $N_{class}$=0 (See Supplementary Note 5). The threshold on the value of $N_{class}$ needed for secure communication is reported in dashed line and the shaded area corresponds to $-1\le SNR \le 1 $. Blue lines correspond to the condition of constant quantum SNR, independently of the value of $N_{class}$. In our experimental conditions, classical measurements have been performed after the second pass in the crystal, at Eve's location (See  Figure \ref{figsetup}). To replicate the detection conditions faced by an eavesdropper lacking access to phase information, we choose a portion of the optical beam where the overlap conditions for detecting interferences are visible. Despite this, the number of detected counts remains consistent with that of the interfering part of the beam. The impact on the classical measurements of the jamming laser is significant, resulting in communication being drowned in noise once $\frac{N_{Class}}{N_{Quantum}}\approx 100$. Quantum measurements on the contrary are unaffected by the jamming laser over 5 order of magnitudes, limited by the available laser power in the signal mode, providing the key tool for secure communication (See equation \ref{eq:fundamental}). We would like to point out that, unlike previous research that studied sensing of undetected photons robustness against external noise \cite{doi:10.1126/sciadv.abl4301,doi:10.1126/sciadv.adg9573,Qian_2023}, our present work focus on the contribution of photons that exhibit the same characteristics as those of the quantum pairs.
Higher signal to noise ratio can be obtained in a communication scheme based on the phase control: in this case a factor two is gained; however this scheme is not suitable for comparison with classical measurement since the phase information cannot be read in the latter configuration (See Supplementary Note 2). Quantum measurements are compared with the theoretical assumption of independence on the number of classical photons (full blue lines),
\begin{figure} 
\includegraphics[scale=0.55]{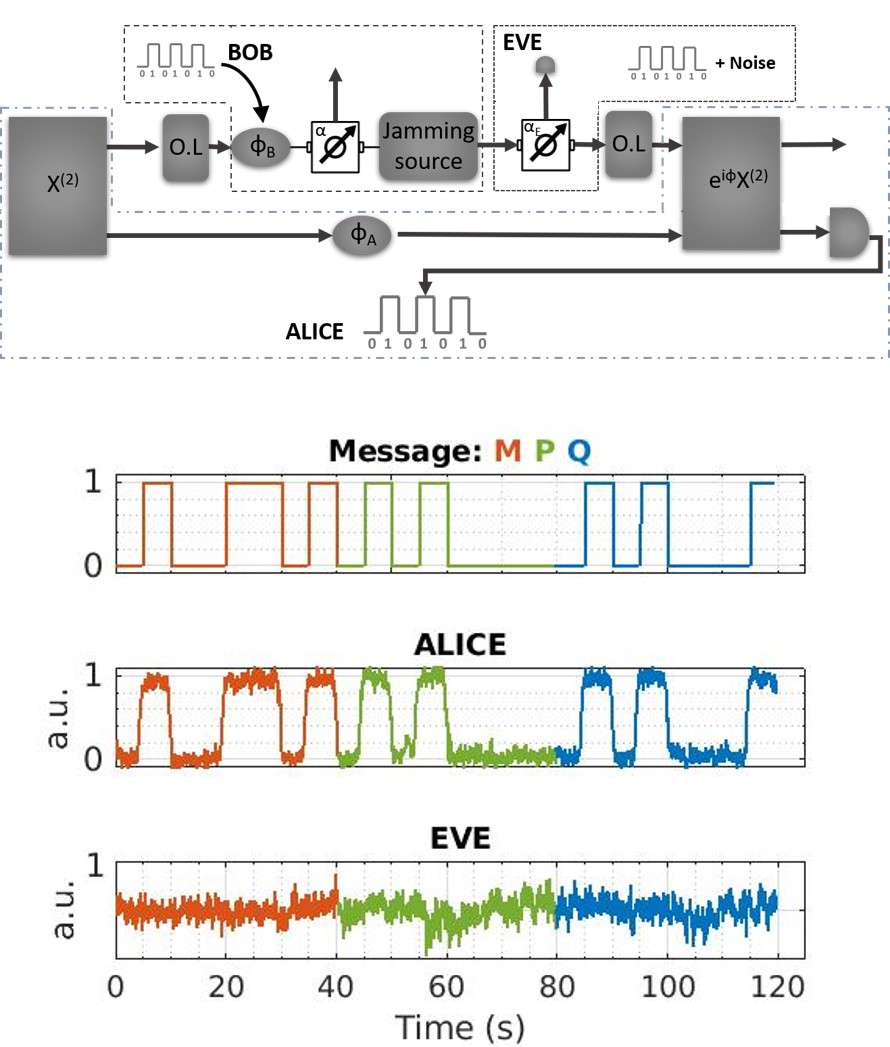}
\centering 
\caption{Experimental demonstration of the transmission of the message "MPQ" encoded in binary. Data are compared for detection at Alice's and Eve's place in the presence of the jamming laser ($\frac{N_{Class}}{N_{Quantum}}\approx 10^5$). The experimental scheme corresponds to the one presented on Figure \ref{figsetup} B.}
\label{figmpq}
\end{figure}
We demonstrate amplitude based secure communication of a three characters message "MPQ" encoded in binary and the results are presented in Figure \ref{figmpq} (The label MPQ corresponds to one of the authors affiliations). We plot on the same time scale the reference message (top), the message read by Alice (central) and the one read by Eve (bottom). While Alice can clearly read the message, Eve is unable to have access to the information. The oscillations in counts measured by Eve comes mainly by the laser instability. 
\begin{figure} 
\includegraphics[scale=0.35]{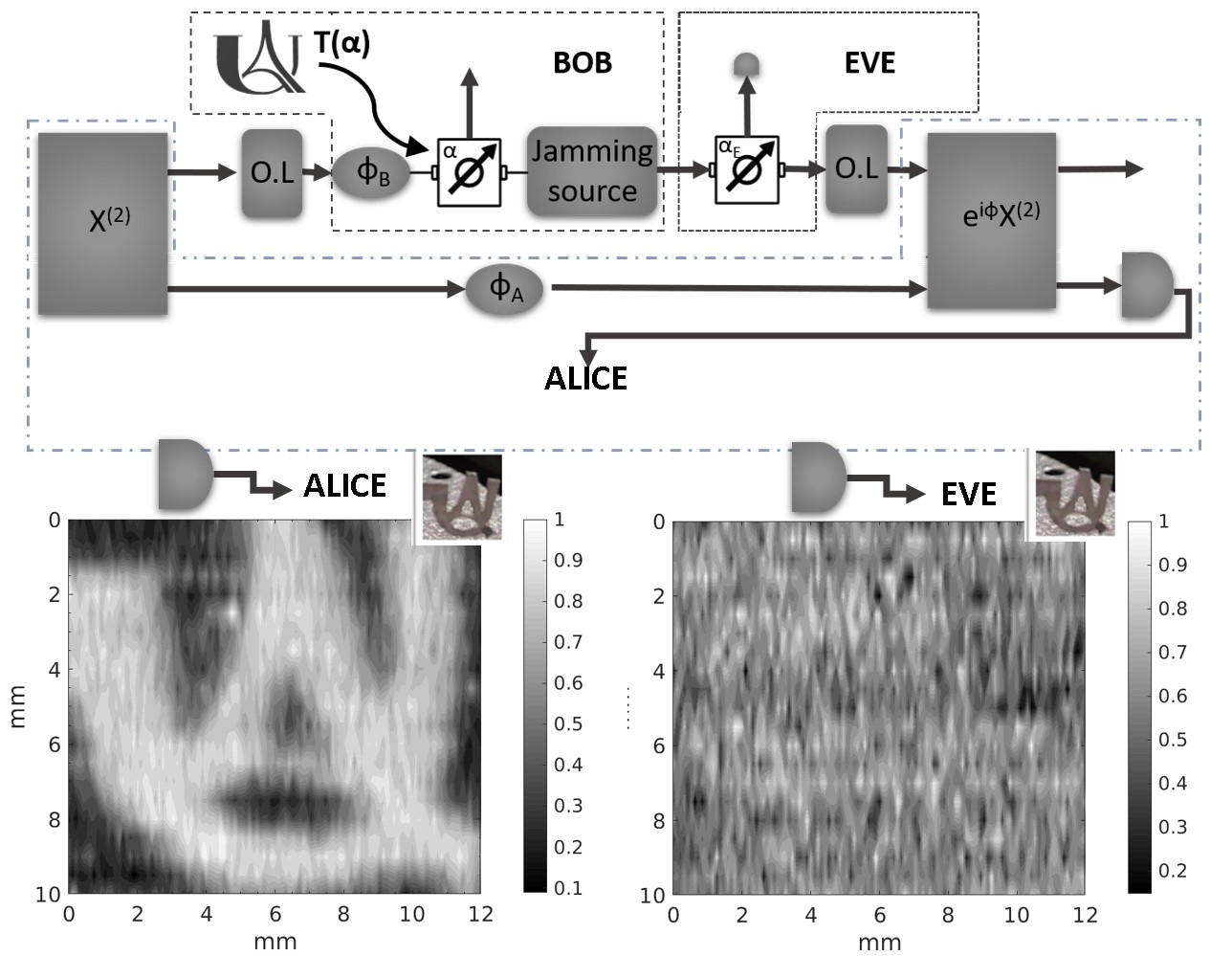}
\centering 
\caption{Experimental demonstration of the transmission of the image of the logo of the University Paris Cité. Data are compared for detection at Alice's and Eve's place in the presence of the jamming laser.The experimental scheme corresponds to the one presented on Figure \ref{figsetup}B}
\label{figlogo}
\end{figure}
The secure communication protocol is also adapted to the transfer of more complex messages such as images, using single pixel detectors. The experimental set-up is presented in Figure \ref{figsetup}B (See Methods). Results are presented in Figure \ref{figlogo} left panel and they are compared to the measurements performed by the eavesdropper (right panel). The object of interest is an opaque foil shaped to the logo of the Université Paris Cité (See the inset). This shape can be clearly recognized by Alice while Eve is completely blinded by the jamming laser. In order to estimate the security of the image transfer we calculated the correlations between measurements at the eavesdropper place and the Alice ones and we found a value of 0.038, comparable to the one that would be obtained for a random distribution of intensity between 0 and 1 (See Supplementary Note 6). 
In the final part of the experimental characterization of the communication protocol we are interested in its speed potential. We use an electrically driven wave plate based on liquid crystal (See Figure \ref{figsetup} C). The optical phase is controlled  with an external bias at a given rate (See Methods). Experimental results are reported in the format of eye diagrams, as conventionally done in classical communication (See Figure \ref{figeye}). An eye shape is obtained whose opening gives information on the communication quality. The horizontal opening of the eye is linked to the time over which the message can be successfully transferred while the vertical width  to the amount of noise that can be tolerated by the communication. The signal slope indicates the sensitivity to timing error; its value has to stay high in robust communication. The experimental results show that the vertical opening stays at around 30$\%$ except for the 8 bit/s where the opening is 15$\%$. The horizontal opening decreases from 60$\%$ to 8$\%$ going from 1.2 to 8 bit/s. The time to pass from the 0 to the 1 level, normalized by half the transmission time of a bit goes from 6$\%$ to 40$\%$ going from 1.2 to 8 bit/s. 

\begin{figure} 
\includegraphics[scale=0.35]{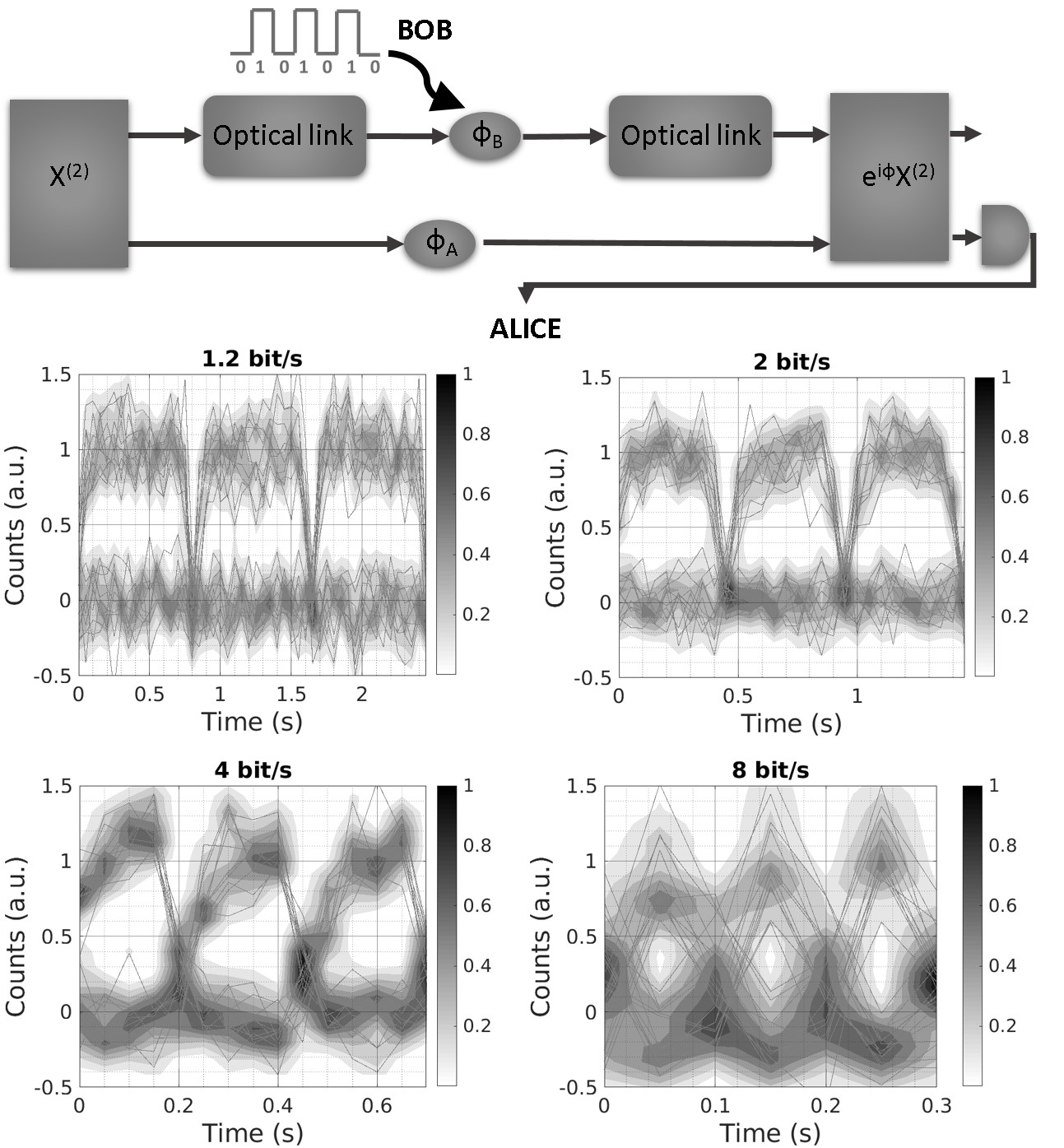}
\centering 
\caption{Experimental demonstration of the transmission of a sequence of 60 bits with rate going from 1.2 bit/s to 8 bit/s. Data are reported as eye diagram. Using the clock signal, the measured counts corresponding to bits 1 0 1 are overlapped on the same time scale with the successive 0 1 0 bits. The procedure is repeated 20 times.}
\label{figeye}
\end{figure}

\section{Discussion}
In this paper, we present a secure communication protocol based on undetected photon sensing. Leveraging the correlation in entangled photon pairs, we employ a technique that exploits this correlation to ensure information security. A jamming source hides the message in the detection noise for an eavesdropper while it stays readable for the intended user thanks to correlation in the entangled photon pairs. We demonstrated the successful communication based on amplitude and phase encoding, achieving transmission rates of up to 8 bit/s and securely transferring a full image between two parties. Moreover, the presented scheme can be easily adapted to fiber-based communication systems. Additionally, the presence of the jamming source serves as a securing tool, which can be exploited to provide a reference for interferometric measurements between the two parties. Faster data exchange can be achieved by increasing the detection efficiency with improved collection efficiency and fiber coupling.
\section{Methods}
\subsection{SNR}
In Figure \ref{figSNR} we report as figure of merit of the communication protocol the photon counts when Bob hides ($C_0$) or does not hide ($C_1$) his path (Figure \ref{figsetup} B) corresponding to the communication levels 0 and 1; here $\Delta\phi=0$. 
SNR=$\frac{C_1-C_0}{\sigma(C_1)}$ where $\sigma(C_1)$ is the standard deviation of the counts distribution measured in the case of Bob not hiding his path (Typical experimental data are presented in Supplementary Note 5). We work in the experimental conditions corresponding the remote possibility of Eve intercepting almost the totality of the message. In order to quantify the number of photons coming from the jamming laser in the same mode of the signal photons we employ single mode fibers. In Figure \ref{figSNR} we report phase based communication measurements in the same experimental conditions. In this case $C_0$ corresponds to $\Delta\phi= \pi$ and $C_1$ to $\Delta\phi=0$ (Where $\Delta\phi=\phi-\phi_A-\phi_B$).
\subsection{Remotely encoded message and image transfer}
Results in Figure \ref{figmpq} are obtained in the following conditions. The experimental set-up is presented in Figure \ref{figsetup}B. Bob writes the message at $\Delta\phi=0$ and he uses a opaque object placed on a remotely controlled translation stage in order to encode the level 0 and 1. The jamming laser is set at $\frac{N_{Class}}{N_{Quantum}}\approx 10^5$.
Based on the same principle, Bob can share in a secure way a two dimensional image. Bob chooses $\Delta\phi=0$ and he horizontally scans the object to image on the optical path (speed 1 mm/s ) and he repeats the procedure for different vertical position. The time constant of the measurements is 50 ms. In this condition we have 10 points per 0.5 millimeter on the horizontal axis; on the vertical axis we have 1 point per 0.5 mm. Alice continuously measures counts on the idler detectors retrieving a bi-dimensional matrix, corresponding to the transmission function of the object. Eve performs the same measurements on the signal detector as she would intercept almost the totality of the message.
\subsection{Eye diagram}
In Figure \ref{figeye} we present phase based communication. The experimental set-up is presented in Figure \ref{figsetup}B. A liquid crystal plate has been programmed to switch between the condition $\Delta\phi=0$, $\Delta\phi=\pi$ regularly, with rate between 1.2 bit/s to 8 bit/s . The counts at Alice's place are recorded with an integration time of 50 ms. Measurements are separated in group of 3 bits and overlapped on the same time scale in order to build the eye diagram. Using the clock signal, the measured counts corresponding to bits 1 0 1 are overlapped on the same time scale with the successive 0 1 0 bits. The procedure is repeated 20 times.

\section{Acknowledgments}
The authors acknowledge the support of the French Agence Nationale de la Recherche (ANR), under grant ANR-23-CE47-0005 (project INTERQUO). ANR and CGI (Commissariat à l’Investissement d’Avenir) are gratefully acknowledged for their financial support of this work through Labex SEAM (Science and Engineering for Advanced Materials and devices), ANR-10-LABX-0096 and ANR-18-IDEX-0001. This work has also been supported by the Paris Île-de-France Région in the framework of DIM SIRTEQ through the Project LION. Authors acknowledge Niccolo Chassagneux for graphical support for the Figure 2 of the main paper.

%


\begin{thebibliography}{30}%
\makeatletter
\providecommand \@ifxundefined [1]{%
 \@ifx{#1\undefined}
}%
\providecommand \@ifnum [1]{%
 \ifnum #1\expandafter \@firstoftwo
 \else \expandafter \@secondoftwo
 \fi
}%
\providecommand \@ifx [1]{%
 \ifx #1\expandafter \@firstoftwo
 \else \expandafter \@secondoftwo
 \fi
}%
\providecommand \natexlab [1]{#1}%
\providecommand \enquote  [1]{``#1''}%
\providecommand \bibnamefont  [1]{#1}%
\providecommand \bibfnamefont [1]{#1}%
\providecommand \citenamefont [1]{#1}%
\providecommand \href@noop [0]{\@secondoftwo}%
\providecommand \href [0]{\begingroup \@sanitize@url \@href}%
\providecommand \@href[1]{\@@startlink{#1}\@@href}%
\providecommand \@@href[1]{\endgroup#1\@@endlink}%
\providecommand \@sanitize@url [0]{\catcode `\\12\catcode `\$12\catcode
  `\&12\catcode `\#12\catcode `\^12\catcode `\_12\catcode `\%12\relax}%
\providecommand \@@startlink[1]{}%
\providecommand \@@endlink[0]{}%
\providecommand \url  [0]{\begingroup\@sanitize@url \@url }%
\providecommand \@url [1]{\endgroup\@href {#1}{\urlprefix }}%
\providecommand \urlprefix  [0]{URL }%
\providecommand \Eprint [0]{\href }%
\providecommand \doibase [0]{http://dx.doi.org/}%
\providecommand \selectlanguage [0]{\@gobble}%
\providecommand \bibinfo  [0]{\@secondoftwo}%
\providecommand \bibfield  [0]{\@secondoftwo}%
\providecommand \translation [1]{[#1]}%
\providecommand \BibitemOpen [0]{}%
\providecommand \bibitemStop [0]{}%
\providecommand \bibitemNoStop [0]{.\EOS\space}%
\providecommand \EOS [0]{\spacefactor3000\relax}%
\providecommand \BibitemShut  [1]{\csname bibitem#1\endcsname}%
\let\auto@bib@innerbib\@empty
\bibitem [{\citenamefont {Wootters}\ and\ \citenamefont
  {Zurek}(1982)}]{Wootters1982}%
  \BibitemOpen
  \bibfield  {author} {\bibinfo {author} {\bibfnamefont {W.~K.}\ \bibnamefont
  {Wootters}}\ and\ \bibinfo {author} {\bibfnamefont {W.~H.}\ \bibnamefont
  {Zurek}},\ }\href {\doibase 10.1038/299802a0} {\bibfield  {journal} {\bibinfo
   {journal} {Nature}\ }\textbf {\bibinfo {volume} {299}},\ \bibinfo {pages}
  {802} (\bibinfo {year} {1982})}\BibitemShut {NoStop}%
\bibitem [{\citenamefont {Bennett}\ \emph {et~al.}(1992)\citenamefont
  {Bennett}, \citenamefont {Bessette}, \citenamefont {Brassard}, \citenamefont
  {Salvail},\ and\ \citenamefont {Smolin}}]{Bennett1992}%
  \BibitemOpen
  \bibfield  {author} {\bibinfo {author} {\bibfnamefont {C.~H.}\ \bibnamefont
  {Bennett}}, \bibinfo {author} {\bibfnamefont {F.}~\bibnamefont {Bessette}},
  \bibinfo {author} {\bibfnamefont {G.}~\bibnamefont {Brassard}}, \bibinfo
  {author} {\bibfnamefont {L.}~\bibnamefont {Salvail}}, \ and\ \bibinfo
  {author} {\bibfnamefont {J.}~\bibnamefont {Smolin}},\ }\href {\doibase
  10.1007/BF00191318} {\bibfield  {journal} {\bibinfo  {journal} {Journal of
  Cryptology}\ }\textbf {\bibinfo {volume} {5}},\ \bibinfo {pages} {3}
  (\bibinfo {year} {1992})}\BibitemShut {NoStop}%
\bibitem [{\citenamefont {{Dieks}}(1982)}]{1982PhLA...92..271D}%
  \BibitemOpen
  \bibfield  {author} {\bibinfo {author} {\bibfnamefont {D.}~\bibnamefont
  {{Dieks}}},\ }\href {\doibase 10.1016/0375-9601(82)90084-6} {\bibfield
  {journal} {\bibinfo  {journal} {Physics Letters A}\ }\textbf {\bibinfo
  {volume} {92}},\ \bibinfo {pages} {271} (\bibinfo {year} {1982})}\BibitemShut
  {NoStop}%
\bibitem [{\citenamefont {Stucki}\ \emph {et~al.}(2011)\citenamefont {Stucki},
  \citenamefont {Legré}, \citenamefont {Buntschu}, \citenamefont {Clausen},
  \citenamefont {Felber}, \citenamefont {Gisin}, \citenamefont {Henzen},
  \citenamefont {Junod}, \citenamefont {Litzistorf}, \citenamefont {Monbaron},
  \citenamefont {Monat}, \citenamefont {Page}, \citenamefont {Perroud},
  \citenamefont {Ribordy}, \citenamefont {Rochas}, \citenamefont {Robyr},
  \citenamefont {Tavares}, \citenamefont {Thew}, \citenamefont {Trinkler},
  \citenamefont {Ventura}, \citenamefont {Voirol}, \citenamefont {Walenta},\
  and\ \citenamefont {Zbinden}}]{Stucki_2011}%
  \BibitemOpen
  \bibfield  {author} {\bibinfo {author} {\bibfnamefont {D.}~\bibnamefont
  {Stucki}}, \bibinfo {author} {\bibfnamefont {M.}~\bibnamefont {Legré}},
  \bibinfo {author} {\bibfnamefont {F.}~\bibnamefont {Buntschu}}, \bibinfo
  {author} {\bibfnamefont {B.}~\bibnamefont {Clausen}}, \bibinfo {author}
  {\bibfnamefont {N.}~\bibnamefont {Felber}}, \bibinfo {author} {\bibfnamefont
  {N.}~\bibnamefont {Gisin}}, \bibinfo {author} {\bibfnamefont
  {L.}~\bibnamefont {Henzen}}, \bibinfo {author} {\bibfnamefont
  {P.}~\bibnamefont {Junod}}, \bibinfo {author} {\bibfnamefont
  {G.}~\bibnamefont {Litzistorf}}, \bibinfo {author} {\bibfnamefont
  {P.}~\bibnamefont {Monbaron}}, \bibinfo {author} {\bibfnamefont
  {L.}~\bibnamefont {Monat}}, \bibinfo {author} {\bibfnamefont {J.-B.}\
  \bibnamefont {Page}}, \bibinfo {author} {\bibfnamefont {D.}~\bibnamefont
  {Perroud}}, \bibinfo {author} {\bibfnamefont {G.}~\bibnamefont {Ribordy}},
  \bibinfo {author} {\bibfnamefont {A.}~\bibnamefont {Rochas}}, \bibinfo
  {author} {\bibfnamefont {S.}~\bibnamefont {Robyr}}, \bibinfo {author}
  {\bibfnamefont {J.}~\bibnamefont {Tavares}}, \bibinfo {author} {\bibfnamefont
  {R.}~\bibnamefont {Thew}}, \bibinfo {author} {\bibfnamefont {P.}~\bibnamefont
  {Trinkler}}, \bibinfo {author} {\bibfnamefont {S.}~\bibnamefont {Ventura}},
  \bibinfo {author} {\bibfnamefont {R.}~\bibnamefont {Voirol}}, \bibinfo
  {author} {\bibfnamefont {N.}~\bibnamefont {Walenta}}, \ and\ \bibinfo
  {author} {\bibfnamefont {H.}~\bibnamefont {Zbinden}},\ }\href {\doibase
  10.1088/1367-2630/13/12/123001} {\bibfield  {journal} {\bibinfo  {journal}
  {New Journal of Physics}\ }\textbf {\bibinfo {volume} {13}},\ \bibinfo
  {pages} {123001} (\bibinfo {year} {2011})}\BibitemShut {NoStop}%
\bibitem [{\citenamefont {Yang}\ \emph {et~al.}(2021)\citenamefont {Yang},
  \citenamefont {Li}, \citenamefont {Ma}, \citenamefont {Qian}, \citenamefont
  {Zhang}, \citenamefont {Wang}, \citenamefont {Zhang}, \citenamefont {Zhou},
  \citenamefont {Tang}, \citenamefont {Wang}, \citenamefont {Yu}, \citenamefont
  {Zhang},\ and\ \citenamefont {Pan}}]{Yang:21}%
  \BibitemOpen
  \bibfield  {author} {\bibinfo {author} {\bibfnamefont {Y.-H.}\ \bibnamefont
  {Yang}}, \bibinfo {author} {\bibfnamefont {P.-Y.}\ \bibnamefont {Li}},
  \bibinfo {author} {\bibfnamefont {S.-Z.}\ \bibnamefont {Ma}}, \bibinfo
  {author} {\bibfnamefont {X.-C.}\ \bibnamefont {Qian}}, \bibinfo {author}
  {\bibfnamefont {K.-Y.}\ \bibnamefont {Zhang}}, \bibinfo {author}
  {\bibfnamefont {L.-J.}\ \bibnamefont {Wang}}, \bibinfo {author}
  {\bibfnamefont {W.-L.}\ \bibnamefont {Zhang}}, \bibinfo {author}
  {\bibfnamefont {F.}~\bibnamefont {Zhou}}, \bibinfo {author} {\bibfnamefont
  {S.-B.}\ \bibnamefont {Tang}}, \bibinfo {author} {\bibfnamefont {J.-Y.}\
  \bibnamefont {Wang}}, \bibinfo {author} {\bibfnamefont {Y.}~\bibnamefont
  {Yu}}, \bibinfo {author} {\bibfnamefont {Q.}~\bibnamefont {Zhang}}, \ and\
  \bibinfo {author} {\bibfnamefont {J.-W.}\ \bibnamefont {Pan}},\ }\href
  {\doibase 10.1364/OE.432944} {\bibfield  {journal} {\bibinfo  {journal} {Opt.
  Express}\ }\textbf {\bibinfo {volume} {29}},\ \bibinfo {pages} {25859}
  (\bibinfo {year} {2021})}\BibitemShut {NoStop}%
\bibitem [{\citenamefont {Chen}\ \emph {et~al.}(2021)\citenamefont {Chen},
  \citenamefont {Zhang}, \citenamefont {Chen}, \citenamefont {Cai},
  \citenamefont {Liao}, \citenamefont {Zhang}, \citenamefont {Chen},
  \citenamefont {Yin}, \citenamefont {Ren}, \citenamefont {Chen}, \citenamefont
  {Han}, \citenamefont {Yu}, \citenamefont {Liang}, \citenamefont {Zhou},
  \citenamefont {Yuan}, \citenamefont {Zhao}, \citenamefont {Wang},
  \citenamefont {Jiang}, \citenamefont {Zhang}, \citenamefont {Liu},
  \citenamefont {Li}, \citenamefont {Shen}, \citenamefont {Cao}, \citenamefont
  {Lu}, \citenamefont {Shu}, \citenamefont {Wang}, \citenamefont {Li},
  \citenamefont {Liu}, \citenamefont {Xu}, \citenamefont {Wang}, \citenamefont
  {Peng},\ and\ \citenamefont {Pan}}]{Chen2021}%
  \BibitemOpen
  \bibfield  {author} {\bibinfo {author} {\bibfnamefont {Y.-A.}\ \bibnamefont
  {Chen}}, \bibinfo {author} {\bibfnamefont {Q.}~\bibnamefont {Zhang}},
  \bibinfo {author} {\bibfnamefont {T.-Y.}\ \bibnamefont {Chen}}, \bibinfo
  {author} {\bibfnamefont {W.-Q.}\ \bibnamefont {Cai}}, \bibinfo {author}
  {\bibfnamefont {S.-K.}\ \bibnamefont {Liao}}, \bibinfo {author}
  {\bibfnamefont {J.}~\bibnamefont {Zhang}}, \bibinfo {author} {\bibfnamefont
  {K.}~\bibnamefont {Chen}}, \bibinfo {author} {\bibfnamefont {J.}~\bibnamefont
  {Yin}}, \bibinfo {author} {\bibfnamefont {J.-G.}\ \bibnamefont {Ren}},
  \bibinfo {author} {\bibfnamefont {Z.}~\bibnamefont {Chen}}, \bibinfo {author}
  {\bibfnamefont {S.-L.}\ \bibnamefont {Han}}, \bibinfo {author} {\bibfnamefont
  {Q.}~\bibnamefont {Yu}}, \bibinfo {author} {\bibfnamefont {K.}~\bibnamefont
  {Liang}}, \bibinfo {author} {\bibfnamefont {F.}~\bibnamefont {Zhou}},
  \bibinfo {author} {\bibfnamefont {X.}~\bibnamefont {Yuan}}, \bibinfo {author}
  {\bibfnamefont {M.-S.}\ \bibnamefont {Zhao}}, \bibinfo {author}
  {\bibfnamefont {T.-Y.}\ \bibnamefont {Wang}}, \bibinfo {author}
  {\bibfnamefont {X.}~\bibnamefont {Jiang}}, \bibinfo {author} {\bibfnamefont
  {L.}~\bibnamefont {Zhang}}, \bibinfo {author} {\bibfnamefont {W.-Y.}\
  \bibnamefont {Liu}}, \bibinfo {author} {\bibnamefont {Li}}, \bibinfo {author}
  {\bibfnamefont {Q.}~\bibnamefont {Shen}}, \bibinfo {author} {\bibfnamefont
  {Y.}~\bibnamefont {Cao}}, \bibinfo {author} {\bibfnamefont {C.-Y.}\
  \bibnamefont {Lu}}, \bibinfo {author} {\bibfnamefont {R.}~\bibnamefont
  {Shu}}, \bibinfo {author} {\bibfnamefont {J.-Y.}\ \bibnamefont {Wang}},
  \bibinfo {author} {\bibfnamefont {L.}~\bibnamefont {Li}}, \bibinfo {author}
  {\bibfnamefont {N.-L.}\ \bibnamefont {Liu}}, \bibinfo {author} {\bibfnamefont
  {F.}~\bibnamefont {Xu}}, \bibinfo {author} {\bibfnamefont {X.-B.}\
  \bibnamefont {Wang}}, \bibinfo {author} {\bibfnamefont {C.-Z.}\ \bibnamefont
  {Peng}}, \ and\ \bibinfo {author} {\bibfnamefont {J.-W.}\ \bibnamefont
  {Pan}},\ }\href {\doibase 10.1038/s41586-020-03093-8} {\bibfield  {journal}
  {\bibinfo  {journal} {Nature}\ }\textbf {\bibinfo {volume} {589}},\ \bibinfo
  {pages} {214} (\bibinfo {year} {2021})}\BibitemShut {NoStop}%
\bibitem [{\citenamefont {Zapatero}\ \emph {et~al.}(2023)\citenamefont
  {Zapatero}, \citenamefont {van Leent}, \citenamefont {Arnon-Friedman},
  \citenamefont {Liu}, \citenamefont {Zhang}, \citenamefont {Weinfurter},\ and\
  \citenamefont {Curty}}]{Zapatero2023}%
  \BibitemOpen
  \bibfield  {author} {\bibinfo {author} {\bibfnamefont {V.}~\bibnamefont
  {Zapatero}}, \bibinfo {author} {\bibfnamefont {T.}~\bibnamefont {van Leent}},
  \bibinfo {author} {\bibfnamefont {R.}~\bibnamefont {Arnon-Friedman}},
  \bibinfo {author} {\bibfnamefont {W.-Z.}\ \bibnamefont {Liu}}, \bibinfo
  {author} {\bibfnamefont {Q.}~\bibnamefont {Zhang}}, \bibinfo {author}
  {\bibfnamefont {H.}~\bibnamefont {Weinfurter}}, \ and\ \bibinfo {author}
  {\bibfnamefont {M.}~\bibnamefont {Curty}},\ }\href {\doibase
  10.1038/s41534-023-00684-x} {\bibfield  {journal} {\bibinfo  {journal} {npj
  Quantum Information}\ }\textbf {\bibinfo {volume} {9}},\ \bibinfo {pages}
  {10} (\bibinfo {year} {2023})}\BibitemShut {NoStop}%
\bibitem [{\citenamefont {Loudon}(2000)}]{loudon_quantum_2000}%
  \BibitemOpen
  \bibfield  {author} {\bibinfo {author} {\bibfnamefont {R.}~\bibnamefont
  {Loudon}},\ }\href@noop {} {\emph {\bibinfo {title} {The {Quantum} {Theory}
  of {Light}}}},\ \bibinfo {edition} {third edition, third edition}\ ed.\
  (\bibinfo  {publisher} {Oxford University Press},\ \bibinfo {address}
  {Oxford, New York},\ \bibinfo {year} {2000})\BibitemShut {NoStop}%
\bibitem [{\citenamefont {Mandel}\ and\ \citenamefont
  {Wolf}(1995)}]{mandel_wolf_1995}%
  \BibitemOpen
  \bibfield  {author} {\bibinfo {author} {\bibfnamefont {L.}~\bibnamefont
  {Mandel}}\ and\ \bibinfo {author} {\bibfnamefont {E.}~\bibnamefont {Wolf}},\
  }\href {\doibase 10.1017/CBO9781139644105.024} {\emph {\bibinfo {title}
  {Optical Coherence and Quantum Optics}}}\ (\bibinfo  {publisher} {Cambridge
  University Press},\ \bibinfo {year} {1995})\BibitemShut {NoStop}%
\bibitem [{\citenamefont {Taleby~Ahvanooey}\ \emph {et~al.}(2019)\citenamefont
  {Taleby~Ahvanooey}, \citenamefont {Li}, \citenamefont {Hou}, \citenamefont
  {Rajput},\ and\ \citenamefont {Chen}}]{e21040355}%
  \BibitemOpen
  \bibfield  {author} {\bibinfo {author} {\bibfnamefont {M.}~\bibnamefont
  {Taleby~Ahvanooey}}, \bibinfo {author} {\bibfnamefont {Q.}~\bibnamefont
  {Li}}, \bibinfo {author} {\bibfnamefont {J.}~\bibnamefont {Hou}}, \bibinfo
  {author} {\bibfnamefont {A.~R.}\ \bibnamefont {Rajput}}, \ and\ \bibinfo
  {author} {\bibfnamefont {Y.}~\bibnamefont {Chen}},\ }\href {\doibase
  10.3390/e21040355} {\bibfield  {journal} {\bibinfo  {journal} {Entropy}\
  }\textbf {\bibinfo {volume} {21}} (\bibinfo {year} {2019}),\
  10.3390/e21040355}\BibitemShut {NoStop}%
\bibitem [{\citenamefont {Cachin}(2004)}]{cachin_informationtheoretic_2004}%
  \BibitemOpen
  \bibfield  {author} {\bibinfo {author} {\bibfnamefont {C.}~\bibnamefont
  {Cachin}},\ }\href {\doibase 10.1016/j.ic.2004.02.003} {\bibfield  {journal}
  {\bibinfo  {journal} {Information and Computation}\ }\textbf {\bibinfo
  {volume} {192}},\ \bibinfo {pages} {41} (\bibinfo {year} {2004})}\BibitemShut
  {NoStop}%
\bibitem [{\citenamefont {Wang}\ and\ \citenamefont
  {Moulin}(2008)}]{wang2008perfectly}%
  \BibitemOpen
  \bibfield  {author} {\bibinfo {author} {\bibfnamefont {Y.}~\bibnamefont
  {Wang}}\ and\ \bibinfo {author} {\bibfnamefont {P.}~\bibnamefont {Moulin}},\
  }\href@noop {} {\bibfield  {journal} {\bibinfo  {journal} {IEEE Transactions
  on Information Theory}\ }\textbf {\bibinfo {volume} {54}},\ \bibinfo {pages}
  {2706} (\bibinfo {year} {2008})}\BibitemShut {NoStop}%
\bibitem [{\citenamefont {Min-Allah}\ \emph {et~al.}(2022)\citenamefont
  {Min-Allah}, \citenamefont {Nagy}, \citenamefont {Aljabri}, \citenamefont
  {Alkharraa}, \citenamefont {Alqahtani}, \citenamefont {Alghamdi},
  \citenamefont {Sabri},\ and\ \citenamefont {Alshaikh}}]{app122010294}%
  \BibitemOpen
  \bibfield  {author} {\bibinfo {author} {\bibfnamefont {N.}~\bibnamefont
  {Min-Allah}}, \bibinfo {author} {\bibfnamefont {N.}~\bibnamefont {Nagy}},
  \bibinfo {author} {\bibfnamefont {M.}~\bibnamefont {Aljabri}}, \bibinfo
  {author} {\bibfnamefont {M.}~\bibnamefont {Alkharraa}}, \bibinfo {author}
  {\bibfnamefont {M.}~\bibnamefont {Alqahtani}}, \bibinfo {author}
  {\bibfnamefont {D.}~\bibnamefont {Alghamdi}}, \bibinfo {author}
  {\bibfnamefont {R.}~\bibnamefont {Sabri}}, \ and\ \bibinfo {author}
  {\bibfnamefont {R.}~\bibnamefont {Alshaikh}},\ }\href {\doibase
  10.3390/app122010294} {\bibfield  {journal} {\bibinfo  {journal} {Applied
  Sciences}\ }\textbf {\bibinfo {volume} {12}} (\bibinfo {year} {2022}),\
  10.3390/app122010294}\BibitemShut {NoStop}%
\bibitem [{\citenamefont {Gregory}\ \emph {et~al.}(2020)\citenamefont
  {Gregory}, \citenamefont {Moreau}, \citenamefont {Toninelli},\ and\
  \citenamefont {Padgett}}]{gregory_imaging_2020}%
  \BibitemOpen
  \bibfield  {author} {\bibinfo {author} {\bibfnamefont {T.}~\bibnamefont
  {Gregory}}, \bibinfo {author} {\bibfnamefont {P.-A.}\ \bibnamefont {Moreau}},
  \bibinfo {author} {\bibfnamefont {E.}~\bibnamefont {Toninelli}}, \ and\
  \bibinfo {author} {\bibfnamefont {M.~J.}\ \bibnamefont {Padgett}},\ }\href
  {\doibase 10.1126/sciadv.aay2652} {\bibfield  {journal} {\bibinfo  {journal}
  {Science Advances}\ }\textbf {\bibinfo {volume} {6}},\ \bibinfo {pages}
  {eaay2652} (\bibinfo {year} {2020})}\BibitemShut {NoStop}%
\bibitem [{\citenamefont {Lloyd}(2008)}]{lloyd_enhanced_2008}%
  \BibitemOpen
  \bibfield  {author} {\bibinfo {author} {\bibfnamefont {S.}~\bibnamefont
  {Lloyd}},\ }\href {\doibase 10.1126/science.1160627} {\bibfield  {journal}
  {\bibinfo  {journal} {Science}\ }\textbf {\bibinfo {volume} {321}},\ \bibinfo
  {pages} {1463} (\bibinfo {year} {2008})}\BibitemShut {NoStop}%
\bibitem [{\citenamefont {Clark}\ \emph {et~al.}(2021)\citenamefont {Clark},
  \citenamefont {Chekhova}, \citenamefont {Matthews}, \citenamefont {Rarity},\
  and\ \citenamefont {Oulton}}]{clark_special_2021}%
  \BibitemOpen
  \bibfield  {author} {\bibinfo {author} {\bibfnamefont {A.~S.}\ \bibnamefont
  {Clark}}, \bibinfo {author} {\bibfnamefont {M.}~\bibnamefont {Chekhova}},
  \bibinfo {author} {\bibfnamefont {J.~C.~F.}\ \bibnamefont {Matthews}},
  \bibinfo {author} {\bibfnamefont {J.~G.}\ \bibnamefont {Rarity}}, \ and\
  \bibinfo {author} {\bibfnamefont {R.~F.}\ \bibnamefont {Oulton}},\ }\href
  {\doibase 10.1063/5.0041043} {\bibfield  {journal} {\bibinfo  {journal}
  {Applied Physics Letters}\ }\textbf {\bibinfo {volume} {118}},\ \bibinfo
  {pages} {060401} (\bibinfo {year} {2021})}\BibitemShut {NoStop}%
\bibitem [{\citenamefont {Moreau}\ \emph {et~al.}(2017)\citenamefont {Moreau},
  \citenamefont {Sabines-Chesterking}, \citenamefont {Whittaker}, \citenamefont
  {Joshi}, \citenamefont {Birchall}, \citenamefont {McMillan}, \citenamefont
  {Rarity},\ and\ \citenamefont {Matthews}}]{moreau_demonstrating_2017}%
  \BibitemOpen
  \bibfield  {author} {\bibinfo {author} {\bibfnamefont {P.-A.}\ \bibnamefont
  {Moreau}}, \bibinfo {author} {\bibfnamefont {J.}~\bibnamefont
  {Sabines-Chesterking}}, \bibinfo {author} {\bibfnamefont {R.}~\bibnamefont
  {Whittaker}}, \bibinfo {author} {\bibfnamefont {S.~K.}\ \bibnamefont
  {Joshi}}, \bibinfo {author} {\bibfnamefont {P.~M.}\ \bibnamefont {Birchall}},
  \bibinfo {author} {\bibfnamefont {A.}~\bibnamefont {McMillan}}, \bibinfo
  {author} {\bibfnamefont {J.~G.}\ \bibnamefont {Rarity}}, \ and\ \bibinfo
  {author} {\bibfnamefont {J.~C.~F.}\ \bibnamefont {Matthews}},\ }\href
  {\doibase 10.1038/s41598-017-06545-w} {\bibfield  {journal} {\bibinfo
  {journal} {Scientific Reports}\ }\textbf {\bibinfo {volume} {7}},\ \bibinfo
  {pages} {6256} (\bibinfo {year} {2017})}\BibitemShut {NoStop}%
\bibitem [{\citenamefont {Moreau}\ \emph {et~al.}(2019)\citenamefont {Moreau},
  \citenamefont {Toninelli}, \citenamefont {Gregory},\ and\ \citenamefont
  {Padgett}}]{moreau_imaging_2019}%
  \BibitemOpen
  \bibfield  {author} {\bibinfo {author} {\bibfnamefont {P.-A.}\ \bibnamefont
  {Moreau}}, \bibinfo {author} {\bibfnamefont {E.}~\bibnamefont {Toninelli}},
  \bibinfo {author} {\bibfnamefont {T.}~\bibnamefont {Gregory}}, \ and\
  \bibinfo {author} {\bibfnamefont {M.~J.}\ \bibnamefont {Padgett}},\ }\href
  {\doibase 10.1038/s42254-019-0056-0} {\bibfield  {journal} {\bibinfo
  {journal} {Nature Reviews Physics}\ }\textbf {\bibinfo {volume} {1}},\
  \bibinfo {pages} {367} (\bibinfo {year} {2019})}\BibitemShut {NoStop}%
\bibitem [{\citenamefont {Johnson}\ \emph {et~al.}(2023)\citenamefont
  {Johnson}, \citenamefont {McMillan}, \citenamefont {Frick}, \citenamefont
  {Rarity},\ and\ \citenamefont {Padgett}}]{johnson_hiding_2023}%
  \BibitemOpen
  \bibfield  {author} {\bibinfo {author} {\bibfnamefont {S.}~\bibnamefont
  {Johnson}}, \bibinfo {author} {\bibfnamefont {A.}~\bibnamefont {McMillan}},
  \bibinfo {author} {\bibfnamefont {S.}~\bibnamefont {Frick}}, \bibinfo
  {author} {\bibfnamefont {J.}~\bibnamefont {Rarity}}, \ and\ \bibinfo {author}
  {\bibfnamefont {M.}~\bibnamefont {Padgett}},\ }\href {\doibase
  10.1364/OE.480881} {\bibfield  {journal} {\bibinfo  {journal} {Optics
  Express}\ }\textbf {\bibinfo {volume} {31}},\ \bibinfo {pages} {5290}
  (\bibinfo {year} {2023})}\BibitemShut {NoStop}%
\bibitem [{\citenamefont {Taylor}(1982)}]{Taylor1982}%
  \BibitemOpen
  \bibfield  {author} {\bibinfo {author} {\bibfnamefont {J.~R.}\ \bibnamefont
  {Taylor}},\ }\href@noop {} {\emph {\bibinfo {title} {An introduction to error
  analysis : the study of uncertainties in physical measurements}}},\ Series of
  books in physics\ (\bibinfo  {publisher} {University Science Books Mill
  Valley, Calif.},\ \bibinfo {address} {Mill Valley, Calif.},\ \bibinfo {year}
  {1982})\BibitemShut {NoStop}%
\bibitem [{\citenamefont {FOX}(2006)}]{FOX_Book}%
  \BibitemOpen
  \bibfield  {author} {\bibinfo {author} {\bibfnamefont {M.}~\bibnamefont
  {FOX}},\ }\href@noop {} {\emph {\bibinfo {title} {Quantum Optics}}},\ edited
  by\ \bibinfo {editor} {\bibfnamefont {O.~M. S.~I.}\ \bibnamefont {PHYSICS}}\
  (\bibinfo  {publisher} {Oxford Master Series in Physics},\ \bibinfo {year}
  {2006})\BibitemShut {NoStop}%
\bibitem [{\citenamefont {Shapiro}\ \emph {et~al.}(2015)\citenamefont
  {Shapiro}, \citenamefont {Venkatraman},\ and\ \citenamefont
  {Wong}}]{shapiro_classical_2015}%
  \BibitemOpen
  \bibfield  {author} {\bibinfo {author} {\bibfnamefont {J.~H.}\ \bibnamefont
  {Shapiro}}, \bibinfo {author} {\bibfnamefont {D.}~\bibnamefont
  {Venkatraman}}, \ and\ \bibinfo {author} {\bibfnamefont {F.~N.~C.}\
  \bibnamefont {Wong}},\ }\href {\doibase 10.1038/srep10329} {\bibfield
  {journal} {\bibinfo  {journal} {Scientific Reports}\ }\textbf {\bibinfo
  {volume} {5}},\ \bibinfo {pages} {10329} (\bibinfo {year}
  {2015})}\BibitemShut {NoStop}%
\bibitem [{\citenamefont {Cardoso}\ \emph {et~al.}(2018)\citenamefont
  {Cardoso}, \citenamefont {Berruezo}, \citenamefont {Ávila}, \citenamefont
  {Lemos}, \citenamefont {Pimenta}, \citenamefont {Monken}, \citenamefont
  {Saldanha},\ and\ \citenamefont {Pádua}}]{cardoso_classical_2018}%
  \BibitemOpen
  \bibfield  {author} {\bibinfo {author} {\bibfnamefont {A.~C.}\ \bibnamefont
  {Cardoso}}, \bibinfo {author} {\bibfnamefont {L.~P.}\ \bibnamefont
  {Berruezo}}, \bibinfo {author} {\bibfnamefont {D.~F.}\ \bibnamefont
  {Ávila}}, \bibinfo {author} {\bibfnamefont {G.~B.}\ \bibnamefont {Lemos}},
  \bibinfo {author} {\bibfnamefont {W.~M.}\ \bibnamefont {Pimenta}}, \bibinfo
  {author} {\bibfnamefont {C.~H.}\ \bibnamefont {Monken}}, \bibinfo {author}
  {\bibfnamefont {P.~L.}\ \bibnamefont {Saldanha}}, \ and\ \bibinfo {author}
  {\bibfnamefont {S.}~\bibnamefont {Pádua}},\ }\href {\doibase
  10.1103/PhysRevA.97.033827} {\bibfield  {journal} {\bibinfo  {journal} {Phys.
  Rev. A}\ }\textbf {\bibinfo {volume} {97}},\ \bibinfo {pages} {033827}
  (\bibinfo {year} {2018})}\ \bibinfo {note} \BibitemShut {NoStop}%
\bibitem [{\citenamefont {Zou}\ \emph {et~al.}(1991)\citenamefont {Zou},
  \citenamefont {Wang},\ and\ \citenamefont {Mandel}}]{PhysRevLett.67.318}%
  \BibitemOpen
  \bibfield  {author} {\bibinfo {author} {\bibfnamefont {X.~Y.}\ \bibnamefont
  {Zou}}, \bibinfo {author} {\bibfnamefont {L.~J.}\ \bibnamefont {Wang}}, \
  and\ \bibinfo {author} {\bibfnamefont {L.}~\bibnamefont {Mandel}},\ }\href
  {\doibase 10.1103/PhysRevLett.67.318} {\bibfield  {journal} {\bibinfo
  {journal} {Phys. Rev. Lett.}\ }\textbf {\bibinfo {volume} {67}},\ \bibinfo
  {pages} {318} (\bibinfo {year} {1991})}\BibitemShut {NoStop}%
\bibitem [{\citenamefont {Wang}\ \emph {et~al.}(1991)\citenamefont {Wang},
  \citenamefont {Zou},\ and\ \citenamefont {Mandel}}]{PhysRevA.44.4614}%
  \BibitemOpen
  \bibfield  {author} {\bibinfo {author} {\bibfnamefont {L.~J.}\ \bibnamefont
  {Wang}}, \bibinfo {author} {\bibfnamefont {X.~Y.}\ \bibnamefont {Zou}}, \
  and\ \bibinfo {author} {\bibfnamefont {L.}~\bibnamefont {Mandel}},\ }\href
  {\doibase 10.1103/PhysRevA.44.4614} {\bibfield  {journal} {\bibinfo
  {journal} {Phys. Rev. A}\ }\textbf {\bibinfo {volume} {44}},\ \bibinfo
  {pages} {4614} (\bibinfo {year} {1991})}\BibitemShut {NoStop}%
\bibitem [{\citenamefont {Lemos}\ \emph {et~al.}(2014)\citenamefont {Lemos},
  \citenamefont {Borish}, \citenamefont {Cole}, \citenamefont {Ramelow},
  \citenamefont {Lapkiewicz},\ and\ \citenamefont {Zeilinger}}]{Lemos2014}%
  \BibitemOpen
  \bibfield  {author} {\bibinfo {author} {\bibfnamefont {G.~B.}\ \bibnamefont
  {Lemos}}, \bibinfo {author} {\bibfnamefont {V.}~\bibnamefont {Borish}},
  \bibinfo {author} {\bibfnamefont {G.~D.}\ \bibnamefont {Cole}}, \bibinfo
  {author} {\bibfnamefont {S.}~\bibnamefont {Ramelow}}, \bibinfo {author}
  {\bibfnamefont {R.}~\bibnamefont {Lapkiewicz}}, \ and\ \bibinfo {author}
  {\bibfnamefont {A.}~\bibnamefont {Zeilinger}},\ }\href {\doibase
  10.1038/nature13586} {\bibfield  {journal} {\bibinfo  {journal} {Nature}\
  }\textbf {\bibinfo {volume} {512}},\ \bibinfo {pages} {409} (\bibinfo {year}
  {2014})}\BibitemShut {NoStop}%
\bibitem [{\citenamefont {Kutas}\ \emph {et~al.}(2020)\citenamefont {Kutas},
  \citenamefont {Haase}, \citenamefont {Bickert}, \citenamefont {Riexinger},
  \citenamefont {Molter},\ and\ \citenamefont {von Freymann}}]{Kutaseaaz8065}%
  \BibitemOpen
  \bibfield  {author} {\bibinfo {author} {\bibfnamefont {M.}~\bibnamefont
  {Kutas}}, \bibinfo {author} {\bibfnamefont {B.}~\bibnamefont {Haase}},
  \bibinfo {author} {\bibfnamefont {P.}~\bibnamefont {Bickert}}, \bibinfo
  {author} {\bibfnamefont {F.}~\bibnamefont {Riexinger}}, \bibinfo {author}
  {\bibfnamefont {D.}~\bibnamefont {Molter}}, \ and\ \bibinfo {author}
  {\bibfnamefont {G.}~\bibnamefont {von Freymann}},\ }\href {\doibase
  10.1126/sciadv.aaz8065} {\bibfield  {journal} {\bibinfo  {journal} {Science
  Advances}\ }\textbf {\bibinfo {volume} {6}} (\bibinfo {year} {2020})}
  \BibitemShut
  {NoStop}%
\bibitem [{\citenamefont {Töpfer}\ \emph {et~al.}(2022)\citenamefont
  {Töpfer}, \citenamefont {Basset}, \citenamefont {Fuenzalida}, \citenamefont
  {Steinlechner}, \citenamefont {Torres},\ and\ \citenamefont
  {Gräfe}}]{doi:10.1126/sciadv.abl4301}%
  \BibitemOpen
  \bibfield  {author} {\bibinfo {author} {\bibfnamefont {S.}~\bibnamefont
  {Töpfer}}, \bibinfo {author} {\bibfnamefont {M.~G.}\ \bibnamefont {Basset}},
  \bibinfo {author} {\bibfnamefont {J.}~\bibnamefont {Fuenzalida}}, \bibinfo
  {author} {\bibfnamefont {F.}~\bibnamefont {Steinlechner}}, \bibinfo {author}
  {\bibfnamefont {J.~P.}\ \bibnamefont {Torres}}, \ and\ \bibinfo {author}
  {\bibfnamefont {M.}~\bibnamefont {Gräfe}},\ }\href {\doibase
  10.1126/sciadv.abl4301} {\bibfield  {journal} {\bibinfo  {journal} {Science
  Advances}\ }\textbf {\bibinfo {volume} {8}},\ \bibinfo {pages} {eabl4301}
  (\bibinfo {year} {2022})}.\ \Eprint
  {http://arxiv.org/abs/https://www.science.org/doi/pdf/10.1126/sciadv.abl4301}
  \BibitemShut
%
\bibitem [{\citenamefont {Fuenzalida}\ \emph {et~al.}(2023)\citenamefont
  {Fuenzalida}, \citenamefont {Basset}, \citenamefont {Töpfer}, \citenamefont
  {Torres},\ and\ \citenamefont {Gräfe}}]{doi:10.1126/sciadv.adg9573}%
  \BibitemOpen
  \bibfield  {author} {\bibinfo {author} {\bibfnamefont {J.}~\bibnamefont
  {Fuenzalida}}, \bibinfo {author} {\bibfnamefont {M.~G.}\ \bibnamefont
  {Basset}}, \bibinfo {author} {\bibfnamefont {S.}~\bibnamefont {Töpfer}},
  \bibinfo {author} {\bibfnamefont {J.~P.}\ \bibnamefont {Torres}}, \ and\
  \bibinfo {author} {\bibfnamefont {M.}~\bibnamefont {Gräfe}},\ }\href
  {\doibase 10.1126/sciadv.adg9573} {\bibfield  {journal} {\bibinfo  {journal}
  {Science Advances}\ }\textbf {\bibinfo {volume} {9}},\ \bibinfo {pages}
  {eadg9573} (\bibinfo {year} {2023})}.\ \Eprint
  {http://arxiv.org/abs/https://www.science.org/doi/pdf/10.1126/sciadv.adg9573}
   \BibitemShut
%
\bibitem [{\citenamefont {Qian}\ \emph {et~al.}(2023)\citenamefont {Qian},
  \citenamefont {Xu}, \citenamefont {Zhu}, \citenamefont {Xu}, \citenamefont
  {Gao}, \citenamefont {Yakovlev}, \citenamefont {Liu}, \citenamefont {Zhu},\
  and\ \citenamefont {Wang}}]{Qian_2023}%
  \BibitemOpen
  \bibfield  {author} {\bibinfo {author} {\bibfnamefont {G.}~\bibnamefont
  {Qian}}, \bibinfo {author} {\bibfnamefont {X.}~\bibnamefont {Xu}}, \bibinfo
  {author} {\bibfnamefont {S.-A.}\ \bibnamefont {Zhu}}, \bibinfo {author}
  {\bibfnamefont {C.}~\bibnamefont {Xu}}, \bibinfo {author} {\bibfnamefont
  {F.}~\bibnamefont {Gao}}, \bibinfo {author} {\bibfnamefont {V.}~\bibnamefont
  {Yakovlev}}, \bibinfo {author} {\bibfnamefont {X.}~\bibnamefont {Liu}},
  \bibinfo {author} {\bibfnamefont {S.-Y.}\ \bibnamefont {Zhu}}, \ and\
  \bibinfo {author} {\bibfnamefont {D.-W.}\ \bibnamefont {Wang}},\ }\href
  {\doibase } {\bibfield  {journal} {\bibinfo
  {journal} {Physical Review Letters}\ }\textbf {\bibinfo {volume} {131}}
  (\bibinfo {year} {2023})}\BibitemShut
  {NoStop}%
\end{thebibliography}

\end{document}


\title{Secure communication based on sensing of undetected photons (Supplementary)}

\author{$J.$ $ Sternberg$}
\affiliation{ Laboratoire Mat\'eriaux et Ph\'enom\`enes Quantiques (MPQ), Universit\'e Paris Cité, CNRS-UMR 7162, Paris 75013, France}

\author{$J.$ $ Voisin$}
\affiliation{ Laboratoire Mat\'eriaux et Ph\'enom\`enes Quantiques (MPQ), Universit\'e Paris Cité, CNRS-UMR 7162, Paris 75013, France}

\author{$C.$ $ Roux$}
\affiliation{ Laboratoire Mat\'eriaux et Ph\'enom\`enes Quantiques (MPQ), Universit\'e Paris Cité, CNRS-UMR 7162, Paris 75013, France}

\author{$Y.$ $ Chassagneux$}
\affiliation{ Laboratoire de Physique de l’ENS, Université PSL, CNRS, Sorbonne Université,Université Paris Cité, Paris, France}

\author{$M.I.$ $ Amanti^{*}$}

\affiliation{ Laboratoire Mat\'eriaux et Ph\'enom\`enes Quantiques (MPQ), Universit\'e Paris Cité, CNRS-UMR 7162, Paris 75013, France}

\maketitle
\section{Supplementary note 1 : Sensing of undetected photons}
In this section we present in more details the sensing of undetected photons and the \textbf{main steps to retrieve equation 1 of the main text}.
The pioneer paper is from Wang et al in 1991(X. Y. Zou, L. J. Wang, and L. Mandel, Phys. Rev. Lett. 67, 318) and the theory is developed in L. J. Wang, X. Y. Zou, and L. Mandel, Phys. Rev. A 44, 4614 (1991).
We consider two identical crystals generating photon pairs by non linear interaction with a pump laser beam.  Using the notation of Wang et al., in the single mode case, the pump field at the crystals 1/2 is $E_{p_{1/2}} \propto V_{p_{1/2}}e^{i(\vec{k}_{p_{1/2}}\vec{r}_{p_{1/2}}-\omega_{p}t)}$, where $\vec{r}_{i}$ is the position of the $i_{th}$ crystal, $\vec{k}_{p_{i}}$ is the wave-vector of the pump photon at the $i$ crystal and $\omega_{p}$ is its frequency. V is associated  to the classical pump field and $|V|^2$ is in units of photons per second at the $i$ crystal. The interaction Hamiltonian operator is $H_{in}=g V e^{i(\omega_{s}+\omega_{i}-\omega_{p})t}e^{-i(\vec{k}_{s}+\vec{k}_{i}-\vec{k}_{p})\vec{r}}\adag_{\vec{k}_s}(\omega_{s})\adag_{\vec{k}_i}(\omega_{i})+H.c.$ in the undepleted pump regime and assuming all the quantum states in the vacuum when the pump laser is turned on. The factor $g$ is proportional to the nonlinear susceptibility of the medium. At a given time $t$ we have the quantum state:
\begin{equation}
\begin{aligned}
\ket{\psi} =\ket{vacuum}+\eta_1V_{1}[e^{i(\omega_{s_{1}}+\omega_{i_{1}}-\omega_{p_{1}})t}e^{-i(\vec{k}_{s_{1}}+\vec{k}_{i_{1}}-\vec{k}_{p_{1}})(\vec{r_{1}}-\vec{r_{o}})}\adag_{\vec{k}_{s_{1}}}(\omega_{s_{1}})\adag_{\vec{k}_{i_{1}}}(\omega_{i_{1}})+\\
\eta_2V_2e^{i(\omega_{s_{1}}+\omega_{i_{2}}-\omega_{p_{2}})t}e^{-i(\vec{k}_{s_{2}}+\vec{k}_{i_{2}}-\vec{k}_{p_{2}})(\vec{r_{2}}-\vec{r_{o}})}\adag_{\vec{k}_{s_{2}}}(\omega_{s_{2}})\adag_{\vec{k}_{i_{2}}}(\omega_{i_{2}})]\ket{vacuum}.
\end{aligned}
\end{equation}
\justify
where $\vec{r_{o}}$ is a given origin position, $|\eta_1|^2$ and $|\eta_2|^2$ are the fraction of incident light converted in photon pairs. We neglect the terms with more than two photons. We assume that the first source is at the origin position and we call $e^{-i\vec{k}_{s_{2}}(\vec{r_{2}}-\vec{r_{1}})}\adag_{\vec{k}_{s_{2}}}(\omega_{s_{2}})\ket{vacuum}=\ket{s_{2}}$ and $e^{-i\vec{k}_{i_{2}}(\vec{r_{2}}-\vec{r_{1}})}\adag_{\vec{k}_{i_{2}}}(\omega_{i_{2}})\ket{vacuum}=\ket{i_{2}}$ , using energy conservation ($\omega_p=\omega_s+\omega_i$) the state at the time $t$ can be written as:
\begin{equation}
\ket{\psi} = \ket{vacuum}+\eta_1V_{1}\ket{i_{1}}\ket{s_{1}}+\eta_2V_2e^{i\phi}\ket{i_{2}}\ket{s_{2}},
\label{eq:stateaoutputsources}
\end{equation}
\justify
with $\phi=\vec{k}_{p_{2}}(\vec{r_{2}})$, $\adag_{\vec{k}_{i_{1}}}(\omega_{i_{1}})\ket{vacuum}=\ket{i_{1}}$ and $\adag_{\vec{k}_{s_{1}}}(\omega_{s_{1}})\ket{vacuum}=\ket{s_{1}}$. We use the assumption that the two crystals are identical. In this framework the overlap condition for the mode $\ket{i_1}$ and $\ket{i_2}$ is $\vec{k}_{i_{1}}=\vec{k}_{i_{2}}$ and
\begin{equation}
e^{-i\vec{k}_{i_{1}}(\vec{r_{2}})}\ket{i_{1}}=\ket{i_{2}} ,
\end{equation}
and
\begin{equation}
e^{-i\vec{k}_{s_{1}}(\vec{r_{2}})}\ket{s_{1}}=\ket{s_{2}},
\end{equation}
and the state of equation \ref{eq:stateaoutputsources} becomes:
\begin{equation}
\ket{\psi_{overlap}} = \ket{vacuum}+\eta_1V_{1}\ket{i_{1}}\ket{s_{1}}+\eta_2V_2e^{i\phi}e^{-i\vec{k}_{i_{1}}(\vec{r_{2}})}e^{-i\vec{k}_{s_{1}}(\vec{r_{2}})}\ket{i_{1}}\ket{s_{1}},
\end{equation}
if $\vec{k}_{i_{1}}\cdot(\vec{r_{2}})=\phi_A$, $\vec{k}_{s_{1}}\cdot(\vec{r_{2}})=\phi_B$ , $V_{1}=V_{2}=V_{p}$ and $\eta_1=\eta_2=\eta$:

\begin{equation}
\ket{\psi_{link}} = \ket{vacuum}+ \sqrt{N_{Quantum}}\ket{i_{1}}\ket{s_{1}}(1+e^{i(\phi-\phi_A-\phi_B)}),
\end{equation}
and equation 1 of the main text is retrieved with $N_{Quantum}=|\eta V_{p}|^2$.

We are now interested to present the \textbf{main step to retrieve of equation 2 of the main text}.
We consider the case of the presence of a beam splitter at Bob's location (See Figure 1 of the main text).
We start from equation \ref{eq:stateaoutputsources} and we obtain, with $V_{1}=V_{2}=V_{p}$ and $\eta_1=\eta_2=\eta$:
\begin{equation}
\ket{\psi_{BS}} =\ket{vacuum}+\sqrt{N_{Quantum}}(\sqrt{1-\alpha^2}\ket{i_{1}}\ket{s_{1}}+i\alpha\ket{i_{1}}\ket{s_{B}})+\sqrt{N_{Quantum}}e^{i\phi}\ket{i_{2}}\ket{s_{2}},
\end{equation}
where $\alpha^2$ is the probability of deviation of the mode $s_1$ into $s_2$ at Bob's place.
If overlap is settled:
\begin{equation}
\ket{\psi_{BS_{overlap}}} =  \ket{vacuum}+\sqrt{N_{Quantum}}(\sqrt{1-\alpha^2}\ket{i_{1}}\ket{s_{1}}+i\alpha\ket{i_{1}}\ket{s_{B}})+\sqrt{N_{Quantum}}e^{i(\phi-\phi_A-\phi_B)}\ket{i_{1}}\ket{s_{1}}.
\label{eq:initalstateBob}
\end{equation}
Alice, who measures $i_1$, has a detection rate, in the case of a detection efficiency $\eta_{det}$ including propagation losses:
\begin{equation}
R_{A}=\eta_{det}N_{Quantum}[(\sqrt{1-\alpha^2}+e^{i(\phi-\phi_A-\phi_B)})(\sqrt{1-\alpha^2}+e^{-i(\phi-\phi_A-\phi_B)})+\alpha^2],
\end{equation}
and
\begin{equation}
R_{A}=2\eta_{det}N_{Quantum}[1+ \sqrt{1-\alpha^2} cos(\phi-\phi_B-\phi_A),
\label{eq:rateAlice}
\end{equation}
as presented in the main text (Equation 2).

\section{Supplementary note 2 :Secure communication protocol}
In this section we present the detailed steps of the communication protocol.
From equation \ref{eq:rateAlice}, we see that measurements at Alice's place are affected by Bob's choice of the phase $\phi_B$ and on his action on the signal path at his place. At the moment of establishing the communication, Alice and Bob choose their paths in order to fulfill $\phi_{i}-\phi_{B_{i}}-\phi_{A_{i}}$=0, where the $i$ label refers to initial conditions. In this case Alice will have a maximum in the detection if Bob choose $\phi_B=\phi_{B_{i}}+2n\pi$ while a minimum will be measured for $\phi_B=\phi_{B_{i}}+(2n+1)\pi$. A binary communication system is established between the two users, thanks to quantum correlations in entangled photon pairs. In the same manner a choice of $\alpha=1$, meaning a complete deviation of the $\ket{s_{1}}$, brings a minimum detection rate. On the other side $\alpha=0$ gives a maximal value for the detection rate, dependent on the choice of the phase $\phi_B$.
The presence of a third user in the scheme can be represented as a loss channel on the signal spatial mode (refer to Figure 1 bottom panel in the main text). We model it as a beam splitter that deviates part of the mode $s_{1}$ into another mode,  $s_{E}$,  with a probability $\alpha_E^2$. The quantum state of equation \ref{eq:initalstateBob} becomes:
\begin{equation}
\begin{aligned}
\ket{\psi_{Eve}}=\ket{vacuum}+\sqrt{N_{Quantum}}(\sqrt{1-\alpha^2}\sqrt{1-\alpha_E^2}\ket{i{_1}}\ket{s{_B}}+ i \sqrt{1-\alpha^2}\alpha_E\ket{i{_1}}\ket{s{_E}}+ i \alpha\ket{i{_1}}\ket{s{_1}})+\\
+\sqrt{N_{Quantum}}e^{i(\phi-\phi_B-\phi_A)}\ket{i{_1}}\ket{s{_1}},
\end{aligned}
\label{eq:initalstateBob1}
\end{equation}
and the detection rate at Alice's place in the presence of an eavesdropper for a detection efficiency $\eta_{det}$:
\begin{equation}
R_{A_{E}}=2\eta_{det}N_{Quantum}(1+\sqrt{1-\alpha^2}\sqrt{1-\alpha_E^2} cos(\phi-\phi_B-\phi_A).
\label{eq:rateAlice1}
\end{equation}
Users can detect the presence of an eavesdropper by observing the visibility of the communication's 0 and 1 levels, upon which it relies. Meanwhile, the intruder can easily access the message addressed to Alice by intercepting the signal photons.
To secure the communication protocol against external attacks, we add a jamming light source that generates photons in the same mode $s_1$ and not correlated with the entangled photons. In this scheme  detection at Alice's location is unaffected by the jamming light since she measures exclusively the idler photons. These photons never left Alice's location and they are not correlated with the noise. On the other side the eavesdropper's detection is strongly affected by the noise since he measures the photon population in the $s_E$ mode. 
The detection rate at the eavesdropper's place in the presence of the jamming laser is:
\begin{equation}
R_{Eve}=\eta_{det_E}N_{Quantum} (1-\alpha^2)\alpha_E^2+\alpha_{E}^2\eta_{det_E}N_{Class},
\label{detection}
\end{equation}
where we assume that, since the jamming beam and the signal photons share the same mode $s_E$, they will also experience the same losses and detection efficiency $\eta_{det_{E}}$. $N_{Class}$ is the number of photons per seconds coming form the classical jamming source in the mode $s_1$.
The noise associated to the detection of the eavesdropper $R_{Eve}$ will be given by the sum of the noise associated to the contributions of the message and the jamming source, since the two are uncorrelated. In the case of $\eta_{det_{E}} \ll 1$, the statistics of detection is Poissonian irrespective of the underlying photon statistics. In the case of higher detection efficiency the statistics of detection is dominated by the one of the incident light. If a laser is employed as jamming beam source, a Poissonian statistic is retrieved. In this work we are in the low gain regime where non linear effects seeded by the jamming laser in the second crystal are neglected. It can also be noticed that the use of an incoherent source would relax this condition. A detailed study of this regime goes beyond the scope of the present work.
In order to secure the protocol we are interested in the noise associated with the counts measurements at Eve's location. We consider for instance the amplitude based communication (case B of the Figure 2 of the main text) and we present the expression of the counts associated to the two communication levels.
If $\alpha=0$ (Bob doesn't deviate the signal) detection counts are:
\begin{equation}
C_{{Eve}_{1}}=\eta_{det_E} \alpha_E^2 T_{meas} ({N_{Quantum}} + N_{Class}),
\end{equation}
if $\alpha=1$, the detection counts are:
\begin{equation}
C_{{Eve}_{0}}=\alpha_{E}^2\eta_{det_E}N_{Class}T_{meas}.
\end{equation}
where $T_{meas}$ is the duration of the measurement.
The difference between the communication level  for $\alpha=0$ and $\alpha=1$ is what determines the readability of the message and it is:
\begin{equation}
C_{{comm}}=\eta_{det_E}{N_{Quantum}} \alpha_E^2T_{meas}.
\end{equation}
The noise associated to the difference $C_{comm}$, $\Delta C_{comm}$, is the sum in quadrature of the noise of $C_{{Eve}_{0}}$ and $C_{{Eve}_{1}}$:
\begin{equation}
\Delta C_{Eve1}=\sqrt{\eta_{det_E} \alpha_E^2 T_{meas} ({N_{Quantum}} + N_{Class})}
\end{equation}
\begin{equation}
\Delta C_{Eve0}=\sqrt{\alpha_{E}^2\eta_{det_E}N_{Class}T_{meas}}
\end{equation}

\begin{equation}
\Delta C_{{comm}}=\sqrt{\eta_{det_E} \alpha_E^2 T_{meas} ({N_{Quantum}} + 2N_{Class})}.
\end{equation}
The signal to noise ratio is 
\begin{equation}
SNR=\frac{\Delta C_{{comm}}}{C_{comm}},
\end{equation}

\begin{equation}
SNR=\frac{\sqrt{\eta_{det_E}{N_{Quantum}\alpha_E^2 T_{meas}} }}{\sqrt{1+2\frac{N_{Class}}{N_{Quantum}}}}.
\end{equation}
This expression is presented in the Figure 2 of the main text as the full red line.

\begin{figure}
\includegraphics[trim={0cm 0 0 0}, scale=0.4]{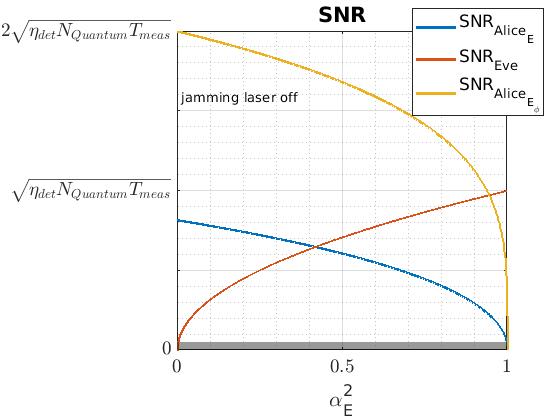}
\centering 
\caption{Calculated SNR in the case of $\eta_{det_E}=\eta_{det}$. Results are reported in the case of the detection performed by Eve (red), by Alice based in the amplitude (blue), by Alice based on the phase (orange). Shaded area corresponds to $0\le SNR \le 1$. Jamming light is off. }
\label{fig:snrvsalphaE}
\end{figure}

If:
\begin{equation}
-1\le SNR \le1
\end{equation}
the message is hidden since the noise overcomes the signal.
This security condition can be simply read as: the shot noise associated to the detection of the jamming light must be higher than the number of quantum photon detected by Eve in the measurement time $T_{meas}$ in order to secure the communication. In more details:
\begin{equation}
\frac{N_{Class}}{N_{Quantum}}>\frac{C_{comm}-1}{2}\approx\frac{C_{comm}}{2}.
\label{thekey}
\end{equation}
This condition corresponds to the shaded red area in the Figure 2 of the main text.
In the Supplementary Note 5 we report the description on the calculation of the secure communication condition from the experimental data.
We report for clarity the calculated SNR for Alice in the case of amplitude based communication. Starting from equation \ref{eq:rateAlice1} the communications counts are 
\begin{equation}
C_{{Comm}_{{Alice}_{E}}}=2\eta_{det}N_{Quantum}T_{meas}\sqrt{1-\alpha_E^2},
\end{equation}
and
\begin{equation}
\Delta C_{{Comm}_{{Alice}_{E}}}=\sqrt{2\eta_{det}N_{Quantum}T_{meas}(2+\sqrt{1-\alpha_E^2})},
\end{equation}
and
\begin{equation}
SNR_{{Alice}_{E}}=\frac{\sqrt{2\eta_{det}N_{Quantum}T_{meas}}\sqrt{1-\alpha_E^2}}{\sqrt{2+\sqrt{1-\alpha_E^2}}} .
\end{equation}
In the case of phase based communication, following the same steps we obtain:
\begin{equation}
SNR_{{Alice}_{E_{\phi}}}=2\sqrt{\eta_{det}N_{Quantum}T_{meas}\sqrt{1-\alpha_E^2}} .
\end{equation}
Assuming in first approximation $\eta_{det_E}=\eta_{det}$ we can compare the SNR for Eve and Alice based on the amplitude and Alice based on the phase, as function of the fraction of the signal stolen by Eve ($\alpha_{E}^2$). Results are reported in Figure \ref{fig:snrvsalphaE}. As presented, the communication protocol is robust against the $\alpha_{E}^2$ value with a slow dependence on this factor. The readability of the message is lost only for $\alpha_{E}^2\approx1$. The jamming laser is off.
·

\newpage
\section{Supplementary note 3: Low gain regime}
In this section we would like to verify the assumption of low gain regime where non linear effects seeded by the jamming laser in the second crystal are neglected. In the case of difference frequency generation at the idler frequency the rate of detected photon is 
\begin{equation}
C_{DFG}=\eta_{det}|\eta \sqrt{V_{p} V_{class}}|^2(1-\alpha_E)(1-\alpha)
\end{equation}
while the rate of detected quantum photons is:
\begin{equation}
C_{Quantum}=\eta_{det}|\eta V_{p}|^2(1-\alpha),
\end{equation}
where $|V_{class}|^2=N_{class}$. In order to obtain $C_{DFG}< C_{Quantum}$ we must have $V_{class}<V_{1}$ meaning that the amount of photons from the jamming source reaching the second source must be inferior to the number of pump photon reaching the same source. In the experimental results presented in this work the pump power is 8 mW corresponding to 2 $\cdot 10^{16}$ photons/s while $N_{classical} \approx 10^5 N_{quantum} $. We measure $N_{Quantum} \eta_{det_{E}}$=4600/s so $N_{classical} \approx 4.6/\eta_{det_{E}} 10^8 $ . Difference frequency generation is then negligible for $\eta_{det_{E}}> 2 \cdot 10^{-8}$ which is a condition in full agreement with usual experimental environment  where $\eta_{det_{E}}$ is estimated to be of the order of $10^{-3}/10^{-4}$.
\newpage
\section{Supplementary note 4: Losses on the communication link}
The visibility of the communication levels $0$ and $1$ is affected by losses and, as in conventional classical link, they determine the ultimate performance. 
A leakage in the communication link, due to optical losses for instance, can be modeled as a beam splitter on the path $\ket{s_2}$, redirecting part of the state to a parasitic channel with an efficiency: $\eta_{losses}^2$. In this scheme the detected signal at Alice's place reads as: 
\begin{equation}
R_{A}=2\eta_{det}N_{Quantum}[1+ \sqrt{1-\eta_{losses}^2} cos(\Delta\phi)],
\end{equation}
where $\Delta\phi=\phi-\phi_{A}-\phi_{B}$.
The visibility of the interference is then affected by the losses.
At the same time a controlled variation of the parameter $\eta_{losses}$ enables a communication scheme based on the amplitude as presented in the main text. Bob can act on $\eta_{losses}$  by deciding the amount of the signal to send back to Alice and his choice is read on the idler photon that never left the laboratory. 

\newpage
\section{Supplementary note 5: Experimental measurements of the SNR in the communication protocol}
\begin{figure} [h]
\includegraphics[trim={0cm 0 0 0}, scale=0.5]{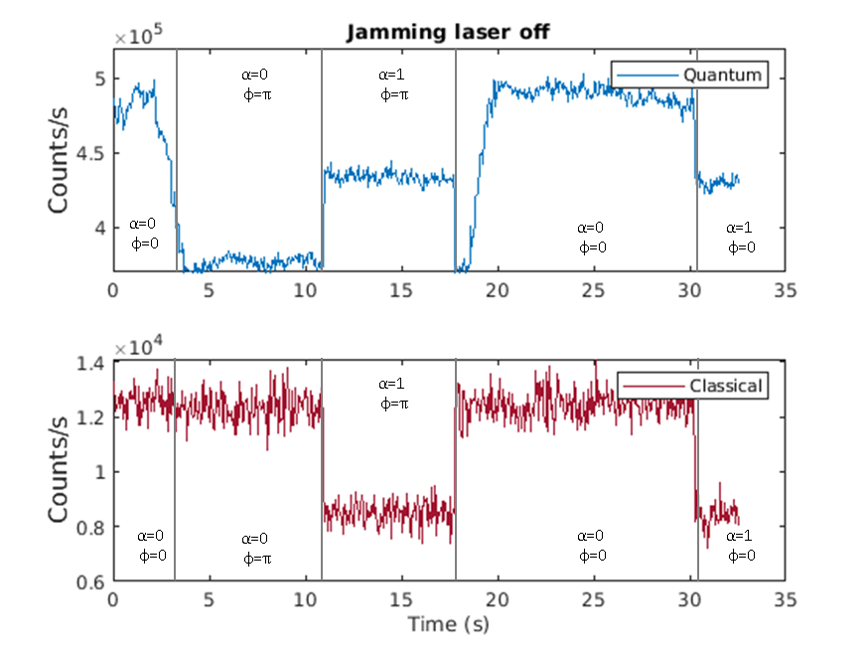}
\centering 
\caption{Experimental Results. Counts as function of the time on Alice's detector (top panel) and Eve's detector (bottom panel). The jamming laser is off.}
\label{fig:laseroff}
\end{figure}

We present here two examples of experimental results that we employed to calculate the values of SNR reported in Figure 2 of the main text. In Figure  \ref{fig:laseroff} we report quantum and classical measurements over time with different values of $\phi_B$ and $\alpha$. 
The blue data correspond to the counts measured on Alice's detector while red ones to Eve's counts. Points are spaced of 50 ms. Starting from the reported traces we calculate the SNR=$\frac{C_1-C_0}{\sigma(C_1)}$: where $\sigma(C_1)$ is the standard deviation of the counts distribution measured in the case of Bob not hiding his path, $C_0$ are the counts when Bob hides ($\alpha=1$) and ($C_1$) are count for $\alpha=0$. Counts have been averaged over 1 s and $\sigma(C_1)$ has been calculated over 20 experimental points. In the error bars we neglect the error on the $\sigma(C_1)$ and the errors on $C_1$ and $C_0$ are the standard deviation associated to the 20 points taken to perform the average.
In order to estimate $N_{Quantum}\eta_{det}\alpha_E^2$  (The measured number of photon pairs per second) we consider the value of counts when the first and second pass are detected ($\alpha=0$) and we subtract the value of counts when Bob hides the signal path ($\alpha=1$). 
In more details (See Figure \ref{fig:laseroff}) $N_{Quantum}$=(12355-8355)1/s= 4000 1/s. Following the results of previous session (See equation \ref{thekey}) we should obtain secure communication for 
\begin{equation}
\frac{N_{Class}} {N_{Quantum}}>N_{Quantum}\eta_{det}\alpha_E^2 T_{meas}/2= 4000 \cdot 0.05/2 =100.
\end{equation}
This value corresponds to the value used in the main text for $C_{Comm}/2$ .

For completeness of the the presentation we show experimental results in the case of the presence of the jamming laser. The laser operating conditions corresponds to the last point of the x scale of the Figure 2 of the main text.
\begin{figure} [h]
\includegraphics[trim={0cm 0 0 0}, scale=0.5]{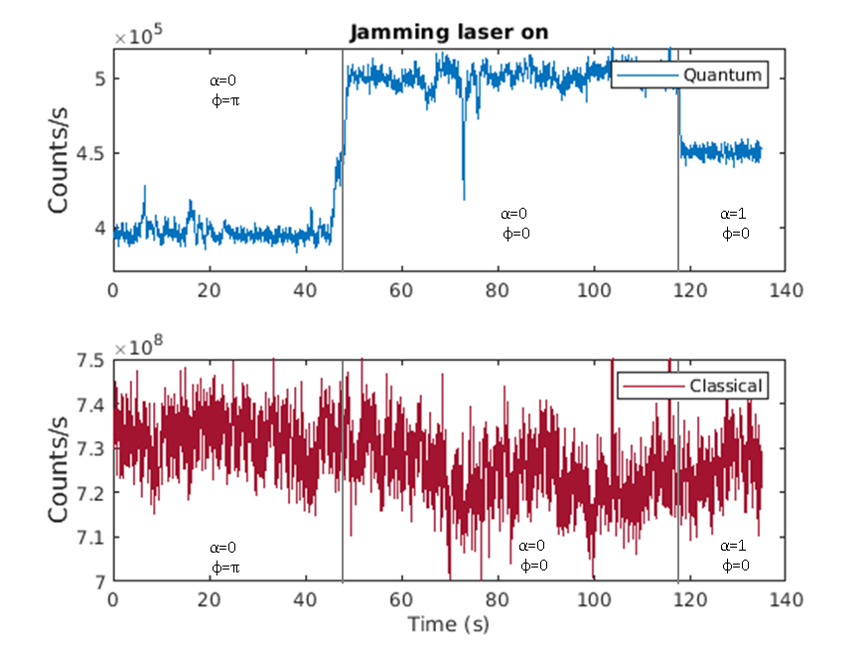}
\caption{Experimental Results. Counts as function of the time on Alice's detector (top panel) and Eve's detector (bottom panel). The jamming laser is on.}
\label{fig:laseron}
\end{figure}

\newpage
\section{Supplementary note 6: Correlation function}
In this section we present the calculated correlation between the image of Figure 5 of the main text and a randomly generated matrix of the same dimension. The calculation has been repeated 100000 times. The results are presented as points on the graphs as function of the iteration number. The red line corresponds to the correlation calculated between our experimental results (Figure 5 the main text). 
\begin{figure}
\includegraphics[trim={0cm 0 0 0}, scale=0.5]{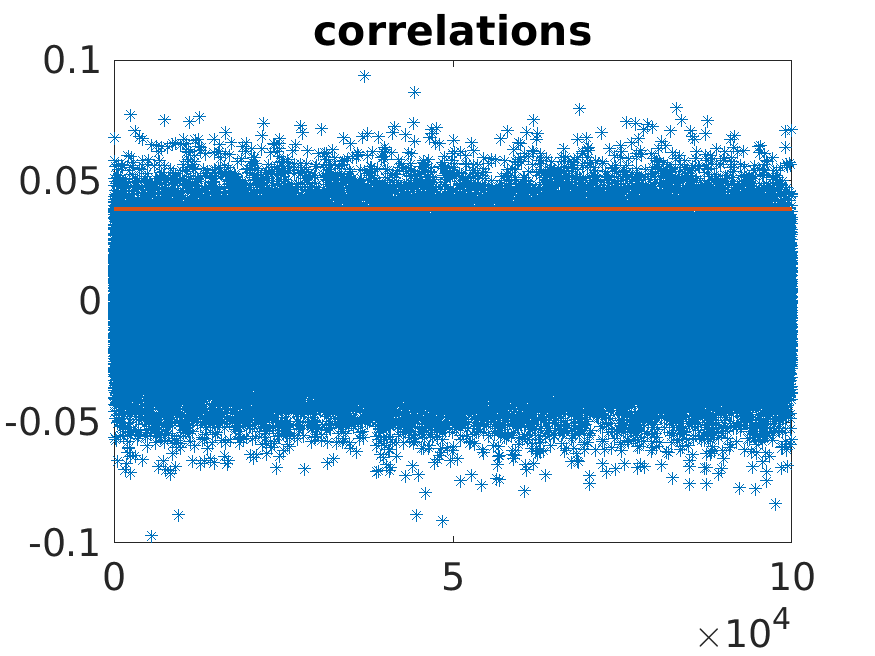}
\caption{Calculated correlations for random generated matrices (blue) and experimental data (red).}
\label{fig:correlation}
\end{figure}